\documentclass[11pt]{article}
\usepackage{amsfonts,amssymb,amsmath,graphics}
\usepackage{geometry,graphicx, color, psfrag}
\usepackage{tikz, tkz-euclide}
\usetikzlibrary{calc}
\usetikzlibrary{decorations.markings}

\usepackage[english]{babel}
\usepackage{color}
\newcommand{\ew}{\color{black}}

\newtheorem{theorem}{Theorem}
\newtheorem{definition}{Definition}
\newtheorem{remark}{Remark}
\newtheorem{proof}{Proof}

\newtheorem{proposition}{Proposition}
\newtheorem{lemma}{Lemma}

\newcommand{\beq}{\begin{eqnarray}}
\newcommand{\eeq}{\end{eqnarray}}

\newcommand{\beqt}{\begin{eqnarray*}}
\newcommand{\eeqt}{\end{eqnarray*}}

\newcommand{\be}{\begin{equation}}
\newcommand{\ee}{\end{equation}}

\newcommand{\bl}{\begin{lemma}}
\newcommand{\el}{\end{lemma}}

\newcommand{\bcon}{\begin{conjecture}}
\newcommand{\econ}{\end{conjecture}}

\newcommand{\br}{\begin{remark}}
\newcommand{\er}{\end{remark}}

\newcommand{\bt}{\begin{theorem}}
\newcommand{\et}{\end{theorem}}

\newcommand{\bd}{\begin{definition}}
\newcommand{\ed}{\end{definition}}

\newcommand{\bp}{\begin{proposition}}
\newcommand{\ep}{\end{proposition}}

\newcommand{\bc}{\begin{corollary}}
\newcommand{\ec}{\end{corollary}}

\newcommand{\bpr}{\begin{proof}}
\newcommand{\epr}{\end{proof}}

\newcommand{\bi}{\begin{itemize}}
\newcommand{\ei}{\end{itemize}}

\newcommand{\ben}{\begin{enumerate}}
\newcommand{\een}{\end{enumerate}}

\newcommand{\Z}{\mathbb Z}

\newcommand{\E}{\mathbb E}

\newcommand{\pee}{\mathbb P}

\newcommand{\s}{\ensuremath{\mathcal{S}}}

\newcommand{\mee}{\ensuremath{\mathcal{M}}}

\begin{document}

\title{{\bf Gibbs Measures for Long-Range Ising Models}}

 \author{ Arnaud Le Ny 
 \ew \footnote{LAMA UPEC $\&$ CNRS, Universit\'e Paris-Est,  94010 Cr\'eteil, France,
 %\newline
 email:  arnaud.le-ny@u-pec.fr.
% \newline 
}
}

\maketitle
{\bf Abstract:} This review-type  paper is based on a talk given at the conference {\em \'Etats de la Recherche en M\'ecanique statistique}, which took place  at IHP in Paris (December 10--14, 2018). We  revisit old results from the 80's about one dimensional long-range polynomially decaying Ising models (often called {\em Dyson models} in dimension one) and describe more recent results about interface fluctuations and interface states in dimensions one  and two.

Based on a series of joint works with R. Bissacot,  L. Coquille, E. Endo, A. van Enter, B. Kimura and W. Ruszel \cite{ELN17, BEELN18, CELNR18, BEEKLNR18, EELN19, ELN19}.

  \tableofcontents

%\vspace{1.5cm}

\newpage
\normalsize

\section{Introduction}

\hspace{.5cm} Gibbs measures for spin systems are probability measures defined on infinite product probability spaces of configurations of spins with values $\pm 1$ attached, in our context, to each site of a lattices $\Z^d$, for $d=1,2,3$ in these notes. They are designed to represent equilibrium states in mathematical statistical mechanics, according to the $2^d$ law of thermodynamics, in the aim of modelling phase transitions and extending Markov chains in a spatial context. 

To avoid uniqueness of probability measures as usually got by the standard Kolmogorov construction in terms of consistent families of marginals at finite volumes, we consider them within the {\em DLR framework}, named after the independent constructions of Dobrushin on one hand \cite{Dob68}, Lanford/Ruelle on the other hand \cite{LR69}, who introduced in the late 60's consistent systems of conditional probabilities w.r.t. the outside of finite volumes. With such use of conditional probabilities and boundary conditions, it appeared indeed  possible to get different probability measures -- thus different global behaviours -- for the same local rules, provided by Gibbs specifications whose task is to specifiy the local conditional probabilities with boundary conditions prescribed outside finite sets. 

The basic example of such spin systems is given by the standard Ising model,  a famous Markov field with a specification given by the standard Boltzmann-Gibbs weights of the form $e^{-\beta H}$ in order to get equilibrium states saturating a variational principle by solving an Entropy-Energy conflict. To get phase transitions, dimension is important, and phase transition in dimension 2 was presented in 1936 by Peierls, followed all over the 20th century by very rich studies on the structure of the convex set of Gibbs measures. In the early 70's, Dobrushin described an even richer structure in higher dimension, with the occurence of rigid interface states in dimension 3, physically stable but non-translation invariant, called Dobrushin states. 

In order to obtain phase transition in dimension 1, Kac/Thompson and Dyson have studied, also  in the late 60's,  infinite range versions of the Ising model, with long-range pair-potentials with polynomial decay leading to phase transition for very slow decays. These probability measures have recently been used to detect interesting phenomenon in  dimension 1, and the extension of such models in dimension 2 for very slow decays had also been recently studied with the hope of interesting interface behaviours not detected in the past.

In these notes, we first describe in Section 2 the DLR framework and standard nearest-neighbours Ising models in dimensions $d=1,2,3$. In Section 3 we focus on long-range Ising models in dimension 1 ({\em Dyson models}), and in Section 4 we describe  recent results for long-range models in dimension 2. 
\section{DLR description of phase transitions -- Ising models on $\mathbb{Z}^d$}

%\subsection{Gibbs measures for general Ising spins}
\hspace{1cm}
We consider {\em Ising spins on  $d$-regular lattices}, {\em i.e.} random variables $\sigma_x,\omega_y,etc.$  attached at each sites $x,y,etc.$ of  $S=\mathbb{Z}^d$ ($d=1,2,3$), and taking  values in the single-spin state-space $E=\{-1,+1\}$. The latter is equipped with the  discrete topology and with the discrete measurable structure, with an {\em a priori} probability counting measure $\rho_0=\frac{1}{2}\big( \delta_{-1} + \delta_{+1} \big)$ and the power set $\mathcal{E}=\mathcal{P}(E)$ as $\sigma$-algebra. We denote by $\mathcal{S}$ the set of all the finite subsets of $S$, and sometimes write $\Lambda \Subset \mathbb{Z}^d$ to denote such a set $\Lambda \in \s$.

 Configurations $\sigma=(\sigma_x)_{x \in S},\omega=(\omega_x)_{x}, etc.$ belong to the {\em Configuration space} $\Omega$,  $$\big( \Omega, \mathcal{F}, \rho  \big) := \big(E^S,\mathcal{E}^{\otimes S},\rho_0^{\otimes S} \big)$$
equipped with the product topology and  measurable structure. For either finite or infinite volumes $\Delta \subset S$, corresponding product spaces will be denoted $(\Omega_\Delta, \mathcal{F}_\Delta, \rho_\Delta)$. We also denote by $\mee_1(\Omega)$ the set of probability measures on them and use the subscript {\em inv} for the restriction to  translation-invariant elements in an obvious sense (see \cite{EFS93}). The set of continuous functions, denoted by $C(\Omega)$, coincide with the set $\mathcal{F}_{\rm qloc}$ of quasilocal functions.

On $(\Omega,\mathcal{F})$, we consider {\em Ferromagnetic pair potentials} $\Phi=\Phi^J$ with coupling functions $$J=(J_{xy})_{x,y \in S},\; J_{xy} \geq 0$$ that are families of local functions $\big(\Phi^J_A\big)_{A \in  \mathcal{S}}$ with $\Phi^J_A=0$ unless $A=\{x,y\}$ for any pair $\{x,y\} \subset S$, in which case for  
any configuration $\sigma \in \Omega$ on has:
\be \label{Dys}\Phi^J_{\{x,y\}} (\sigma)=J_{xy}\sigma_x \sigma_y
\ee

We shall now forget the supscript $J$, and focus on two main types of coupling functions $J$:
\begin{itemize}
\item{\bf Classical (homogeneous) {\em n.n.} Ising model: }

The interaction $\Phi_A$ is non-null only for pairs $A=\{x,y\}$ of nearest-neighbours ({\em n.n.}), also sometimes  briefly written $A=\langle xy \rangle $, with couplings $J=J^{n.n.}$ given by 
% (edge of S.
$$
J_{xy}=J^{n.n.}_{xy}:  = J \mathbf{1}_{|x-y|=1},\; J>0,\; {\rm for \; all} \;  \{x,y\} \subset \mathbb{Z}^d
$$

\item{\bf  Long-range ferromagnetic Ising models with polynomial decay $\alpha > d$:}
\begin{equation}\label{Jalpha}
J_{xy}=J^\alpha_{xy}:= \frac{J}{|x-y|^\alpha}, J>0,\;\; {\rm for \; all} \;  \{x,y\} \subset \mathbb{Z}^d
\end{equation}
where $| \cdot |$ denotes a canonical norm on $\mathbb{Z}^d$, with $\alpha >d$ so that the couplings are summable:
$$
\forall x \in \mathbb{Z}^d, \sum_{y \in \Z^d} |J^\alpha_{xy}| < \infty
$$
\end{itemize}
Given a finite volume $\Lambda$ in $\mathbb{Z}^d$, for a prescribed boundary condition ({\em b.c.}) $\omega_{\Lambda^c} \in \Omega_{\Lambda^c}=\{-1,1\}^{\Lambda^c}$, we define Hamiltonians on $\Omega$ for any $\sigma \in \Omega$ to be\footnote{We  also sometimes consider 'magnetic fields', either homogeneous ($h\in \mathbb{R}$), inhomogeneous ($h=(h_x)_{x\in S}$), or  random ($h=(h_x[\eta])_{x \in S}$ for random variables  $\eta$'s playing the role of disorder in the {\em Random Field Ising Model}). The formal Hamiltonian  reads then 
$
H_\Lambda[\eta] = - \sum_{x,y \in S} J_{xy} \sigma_x \sigma_y - \sum_{x \in S} h_x[\eta] \sigma_x
$.} the uniformly convergent series
\begin{equation}
H^{\omega}_\Lambda(\sigma_{\Lambda}) = - \sum_{\substack{x,y\in \Lambda \\ x\neq y}} J_{xy} \sigma_x \sigma_y - \sum_{\substack{\substack{x \in \Lambda\\ y \in \Lambda^{c}}}} J_{xy} \sigma_x \omega_y
\end{equation}
For a fixed inverse temperature $\beta>0$, the {\em Gibbs specification} is determined by a family of probability kernels $\gamma=(\gamma_\Lambda)_{\Lambda \in \s}$ defined on $\Omega_\Lambda \times \mathcal{F}_{\Lambda^c}$ 
 by the Boltzmann-Gibbs weights
 \be \label{GibbsSpe}
\gamma_\Lambda (\sigma | \omega)= 
\frac{1}{Z_\Lambda^\omega} e^{-\beta H^{\omega}_\Lambda (\sigma_{\Lambda})}
\ee
where $
Z^{\omega}_{\Lambda} = 
\sum_{\sigma\in \Omega_{\Lambda}} e^{-\beta H^{\omega}_\Lambda (\sigma_{\Lambda})}
$ is the {\em partition function},  related to free energy. 
\br Due to this Boltzmann-Gibbs form (\ref{GibbsSpe}), finite-volume Gibbs measures at temperature $T=\beta^{-1} >0$ are designed to maximize {\em Entropy minus Energy}, satisfying a variational principle in concordance with the $2^d$ principle of thermodynamics. After some work,  infinite-volume Gibbs measures are also shown to  represent equilibrium states at infinite volume : they are  the one(s) who {\em minize(s) free energy $"F=U-TS"$}, or equivalently the one(s) that, {\em at a fixed 'energy', maximize(s) 'entropy'}. We do not describe this {\em variational approach} in these notes, although it justifies the heuristics behind {\em Entropy-Energy arguments} used in the low temperature proofs of phase transitions within Peierls or Pirogov-Sinai strategies \cite{Ruelle68, BLP79, EFS93,SS81, LN09}.
\er
\hspace{.5cm} Within this DLR Framework, a {\em Gibbs measure} $\mu$ is then defined to be a probability measure on $\mee_1(\Omega)$ whose conditional probabilities with boundary condition $\omega$ outside $\Lambda$, are of the form of the kernels  $\gamma_\Lambda (\cdot | \omega)$ and thus satisfy the {\em DLR equations}:
%$\mu^\omega_\Lambda$,
%\be \label{DLR1}
% \mu^\omega_\Lambda (\sigma)= 
% \begin{cases}
% \frac{1}{Z_\Lambda^\omega} e^{-\beta H_\Lambda (\sigma | \omega)}, &\text{ if }\sigma_{\Lambda^c}=\omega_{\Lambda^c},\\
% 0, &\text{ otherwise}.
% \end{cases}
% \ee
%whose closed form is 
\be \label{DLR1}
\mu \gamma_\Lambda = \mu, \text{ for all } \Lambda \Subset \Z
\ee
%where $\Lambda \Subset \Z$ means that $\Lambda$ is a finite subset of $\Z$. 
Alternatively, DLR equations (\ref{DLR1}) satisfied by Gibbs measures $\mu$ for a specification $\gamma$ read
\begin{equation}\label{DLR}
\mu (\cdot) = \int \frac{1}{Z_\Lambda^\omega} e^{-\beta H_\Lambda(\cdot_\Lambda \omega_{\Lambda^c})} d\mu(\omega)
\end{equation}
Equation (\ref{DLR}) is the starting point of the  extremal decompositions of Gibbs measures leading to the Choquet simplex structure of sets of Gibbs measures (see below).

DLR equations  (\ref{DLR}) and (\ref{DLR1}) also mean that, for a subset $\Lambda$ {\em finite}, regular versions of conditional probabilities of $\mu$ w.r.t. $\mathcal{F}_{\Lambda^c}$ should satisfy
$$
\mu[\; \cdot \; | \mathcal{F}_{\Lambda^c}][\omega) = \frac{1}{Z_\Lambda^\omega} e^{-\beta H_\Lambda^\omega(\cdot_\Lambda )},\; \mu{\rm - a.s.}(\omega)
$$

We denote by $\mathcal{G}(\gamma)$ the set of Gibbs measures, and $\mathcal{G}_{\rm inv}(\gamma)$ for translation-invariant ones.

Existence of Gibbs measures  ($\mathcal{G}(\gamma) \neq 0$) is insured by our compact finite-state space framework, and more generally from the existence of continuous versions of conditional probabilities (equivalent to {\em Quasilocality}, see \cite{EFS93, Geo88,LN09} or Section \ref{SectionNG}).

 In Equilibrium statistical mechanics, one is  more often  interested  in multiplicity of Gibbs measures, called {\em phase transition} when $|\mathcal{G}| > 1$. In next subsection, we describe the fondamental case of classical Ising models  where two different {\em phases} exist at low temperature, because then  entropic effects cannot perturbate enough energetic minimizers.

 In such cases, a general result\footnote{There is a  proof of this result avoiding  Krein-Milman Theorem abstract theorem. It can be made within a similar scheme as the ergodic decomposition theorem or de Finetti's description of exchangeable measures, following a general demonstration of Dynkin, see \cite{Dynkin78, Geo88, LN09}.}
 on DLR measures is the following:
\begin{theorem}\label{Geo,LN09}\label{pipoint}
The set $\mathcal{G}(\gamma)$ of DLR measures for a given specification $\gamma$
 is a convex subset of
$\mathcal{M}_1^+(\Omega)$
 whose extreme boundary is denoted ${\rm ex}  \mathcal{G}(\gamma)$, and satisfies  the following properties:
\begin{enumerate}
\item The extreme elements of $\mathcal{G}(\gamma)$ are the
probability measures $\mu \in \mathcal{G}(\gamma)$ that are
trivial on the tail $\sigma$-field $\mathcal{F}_\infty := \cap_{\Lambda \in \s} \mathcal{F}_{\Lambda^c}$:
\begin{equation}\label{tailext}
{\rm ex}\mathcal{G}(\gamma)= \Big \{ \mu \in \mathcal{G}(\gamma):
\mu(B)=0 \; {\rm or} \; 1, \; \forall B \in    \mathcal{F}_\infty
\Big \}
\end{equation}
Moreover, distinct extreme elements $\mu,\nu \in {\rm
ex}\mathcal{G}(\gamma)$ are mutually singular: $\exists B \in
\mathcal{F}_\infty$, $\mu(B)=1$ and $\nu(B)=0$, and more
generally,
 each $\mu \in \mathcal{G}(\gamma)$ is uniquely determined within $\mathcal{G}(\gamma)$ by its restriction to $\mathcal{F}_\infty$
\item $\mathcal{G}(\gamma)$ is a  {\em Choquet simplex}: Any $\mu
\in \mathcal{G}(\gamma)$ can be written in a unique way as
\begin{equation}\label{Choquet}
\mu = \int_{{\rm ex}\mathcal{G}(\gamma)} \; \nu \cdot \alpha_\mu(d
\nu)
\end{equation}
where $\alpha_\mu \in \mathcal{M}_1^+\big({\rm
ex}\mathcal{G}(\gamma),e({\rm ex}\mathcal{G}(\gamma))\big)$ is
defined for all $M \in e({\rm ex}\mathcal{G}(\gamma))$ by
(\ref{weights}) below.
%\be \label{weights} \alpha_\mu (M) = \mu \Big[
%\big \{ \omega \in \Omega: \exists \nu \in M,\; \lim_n
%\gamma_{\Lambda_n} (C | \omega) = \nu(C) \;{\rm for \; any \;
%cylinder} \; C \big \} \Big]. \ee
\end{enumerate}
\end{theorem}
The weights $\alpha_\mu (M)$ are associated with any measurable subset of measures
$M \in e({\rm ex}\mathcal{G}(\gamma))$, the $\sigma$-algebra of evaluation maps on spaces of measures \cite{FV16, Geo88, LN09}. They represent the relative weights of typical configurations of the extremal Gibbs measures in the mixture (\ref{Choquet}), \be \label{weights}
\alpha_\mu (M) = \mu \Big[ \big \{ \omega \in \Omega: \exists \nu
\in M,\; \lim_n \gamma_{\Lambda_n} (C | \omega) = \nu(C) \;,
\forall C \in \mathcal{C} \big \} \Big] \ee

Extremal Gibbs measures are sometimes called {\em States} or {\em Phases}, while {\em Pure states} concern translation-invariant extremal Gibbs measures, such as the $+$- or $-$-states  got by weak limits with homogeneous all $+$- or all $-$-boundary conditions in our ferromagnetic spin systems.

We emphasize that even concerning $n.n.$ homogeneous Ising models, there can be infinitely many's non-transition-invariant extremal Gibbs measures entering in the extremal decomposition (\ref{Choquet}), for {\em e.g.} $d=3$ or on Cayley trees \cite{Geo88}. We describe the case of anisotropic long-range models in dimension 2, for which this holds for slow decays of the interaction (Section 4.2).

Except in one occasion (for some decays $\alpha \in (3,4)$, see Section 3), we shall consider mostly {\bf ferromagnetic} couplings {\em i.e.} $J\geq 0$. In particular, we enjoy {\em FKG and monotonicity preserving properties}, among other reasons because they yield the existence of two  extremal infinite-volume Gibbs measures as weak limits of the all $-$- or all $+$-boundary conditions:
\begin{equation} \label{mupm}
\mu^-(\cdot):=\lim_{\Lambda \uparrow \Z^d} \gamma_\Lambda^{\beta \Phi} (\cdot | +)\; {\rm and} \; \mu^+(\cdot):=\lim_{\Lambda \uparrow \Z^d} \gamma_\Lambda^{\beta \Phi} (\cdot | +)
\end{equation}

In this framework, uniqueness is insured by $\mu^- = \mu^+$, while  phase transition is got by proving that  $\mu^- \neq \mu^+$.
Moreover, for any other Gibbs measure $\mu \in \mathcal{G}(\gamma)$, the following stochastic domination inequalities hold:
$$
\mu^- \leq \mu \leq \mu^+ 
$$
Here we use the {\em FKG order} '$\leq$',  meaning that the bounds are valid for expectations of increasing functions. 
We shall sometimes write such expectations $\langle \cdot \rangle^+$,  $\langle \cdot \rangle^-$,  $\langle \cdot \rangle^0$,  and $\langle \cdot \rangle^\pm$, for respectively the all $+$-,  all $-$-,  free,  and $\pm$-"Dobrushin boundary condition", or $\langle \cdot \rangle^\omega$ for general b.c. $\omega$. We add a subscript $\Lambda$, or sometimes $L$, and write  $\langle \cdot \rangle^\cdot_L$ for the finite-volume versions on square boxes $\Lambda=\Lambda_L=([-L,+L] \cap \Z)^d$, and also $\pee_L$ for the corresponding probabilities. 

In the particular case of $n.n.$ Ising models, for which the precise results that $\mathcal{G}(\gamma)=[\mu^-,\mu^+]$ in $2d$ but not in $3d$ have been established in the seventies at low temperature for $n.n.$ Ising models, culminating with the independent results of Aizenman or Higuchi around 1980 \cite{Aiz,Hig}. In Section 4, we  provide hints to prove that the absence of  translation-invariant extremal states other than $\mu^-$ and $\mu^+$ is also true for long-range polynomially decaying potentials in $2d$ (at least for fast decays $\alpha >3$) and provide partial results from \cite{CELNR18} for very long range models with slow decays $\alpha \in (2,3)$.

As another general result for $n.n$ models, let us quote the explicit values of the {\em magnetization} $\mu[\sigma_0]$ in $2d$ by  Onsager (\cite{Ons44}, 1944) and the result of uniqueness in homogeneous  fields by Lee and Yang (\cite{YL}, 1952). For a complete, rigorous and didactic presentation of this classical Ising model, one should really read  the  book of  Friedli/Velenik (\cite{FV16}, Chap. 3).

\subsection{Phase transition {\em vs.} uniqueness results in the classical $n.n.$ case}
\begin{itemize}
\item{\bf Uniqueness in dimension $d=1$}
\end{itemize}

For $n.n.$ Ising models and more generally {\em finite-range random fields in one dimension}, uniqueness is well known due to existence and uniqueness results of the invariant measure of  irreducible Markov chains, see {\em e.g.} Chapter 3 of \cite{Geo88}. Called in generality {\em Markov random fields}, they are indeed also {\em reversible Markov chains} and there is a one to one correspondance : one says that {\em Global and local Markov properties are equivalent}. This is not always the case, as seen in {\em e.g.}  \cite{Foll, Gold2, vW} or to some extend for long range models, see Section 2.3. and \cite{BEELN18}. 

Heuristically, writing the free energy under the form $"F=U-TS"$, one hase two so-called {\em Ground states} (minimizers of the Hamiltonian), the all $+$- and the all $-$-{\em configurations}. Inserting a droplet of defects in one of this phase would have a constant, volume-independent, energetic cost. It is thus always  beaten by entropy in the thermodyamic limit, at any positive temperature. Thus, at any temperature, only a unique disorder phase appears.

 For a more rigorous presentation of such Entropy {\em vs.} Energy arguments, see \cite{Ruelle68,BLP79, SS81}.

% \\

%More general framework of {\em Markov fields}.....Equivalence with Markov chains. (cf local {\em vs.} global Markov field in dim $d=3$.

\begin{itemize}
\item{\bf Phase transition  at low temperature for $d=2$}
\end{itemize}

We shall briefly sketch the standard argument of Peierls to prove phase transition for the $2d$ Ising model at low temperature, but state first a more general result. For the full convex picture at any temperature, with only two translation-invariant extremal Gibbs measures see \cite{Aiz, Hig, Geo88} or the discussion in \cite{CELNR18}.
\begin{theorem}
Consider $\gamma$ to be the specification (\ref{GibbsSpe}) of the $n.n.$ Ising model in dimension 2. Then there exists a critical inverse temperature $0 < \beta_c < + \infty$ such that
\begin{itemize}
\item $\mathcal{G}(\gamma)=\{\mu \}$ for all $\beta < \beta_c$.
\item $\mathcal{G}(\gamma)=[\mu^- , \mu^+]$ for all  $\beta > \beta_c$ where the {\em extremal phases} $\mu^- \neq \mu^+$ can be selected via "$-$"-or "$+$"-boundary conditions: for all $f \in \mathcal{F}_{\rm{qloc}}$,
\be \label{WLimits}
\mu^-[f]:=\lim_{\Lambda \uparrow \mathcal{S}} \gamma_\Lambda [f \mid -] \; {\rm and} \; \mu^+[f]:=\lim_{\Lambda \uparrow \mathcal{S}} \gamma_\Lambda [f \mid +].
\ee
Moreover, for any $\mu \in \mathcal{G}(\gamma)$, for any bounded {\em increasing} $f$, 
$\mu^-[f] \; \leq \; \mu[f] \; \leq \; \mu^+[f]$,
and the extremal phases have opposite {\em magnetizations}\footnote{We denote  
$\mu[f]$ for the expectation $\E_\mu[f]$.} $m^*(\beta):=\mu^+ [\sigma_0]=-\mu^- [\sigma_0]>0$.
\end{itemize}
\end{theorem}
%\large
\begin{figure}[ht]
		\centering
		\includegraphics[width=0.33\textwidth]{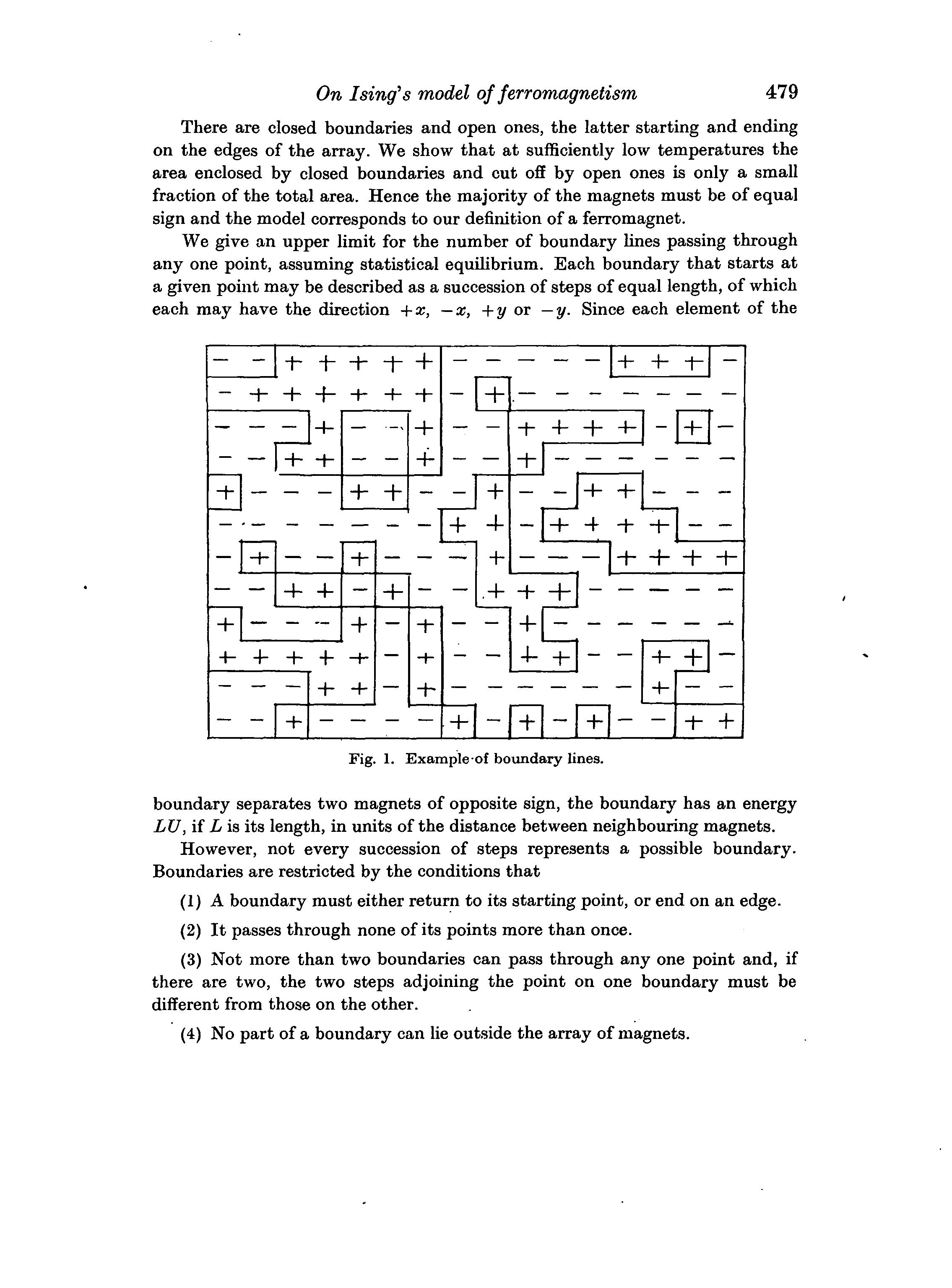}
	\caption{Original Peierls contours}
		\label{fig-original}
\end{figure}
%\normalsize
 Let us sketch now the {\em Peierls argument} (\cite{Peierls36}, 1936), also derived/discovered by Griffiths (\cite{Grif64} (1964)) or Dobrushin (\cite{Dob65}, 1965). This  geometrical approach uses the interface between different spin values $+$'s and $-$'s to define the Hamiltonian in terms of closed circuits called {\em Contours}, to get  temperature-dependent bounds on the energy of  configurations, that eventually leads to phase transition at low temperature by rigorous entropy {\em vs.} energy arguments. For a complete presentation of the argument, one could {\em e.g.} consult \cite{LP2017}.

%{\bw  A {\em contour $\gamma$ of the dual of $\mathbb{Z}^2$ is said to occur} in the configuration $\sigma$, or simply to be a contour {\em of} $\sigma$, if it separates some "$+$" and "$-$" areas of $\omega$, i.e. if $\gamma \subset \big\{ b + (\frac{1}{2},\frac{1}{2}) : b=\{i,j\}, ||i-j||_1=1, \omega_i \neq \omega_j \big\}$. 
%and 

Consider a finite volume $\Lambda \in \mathcal{S}$, start with the boundary condition $+$  and take the probability measure $\gamma_\Lambda(\cdot \mid +)$.
A  {\em path} -- in $\Z^2$ is a finite sequence $\pi=\{i_1,\dots,i_n\}$ of sites such that $i_j$ and $i_k$ are $n.n.$ ( $|j-k|=1$).  We call dual of $\Z^2$ the set $\Z^2 + (\frac{1}{2}, \frac{1}{2})$ and define a {\em contour} $\gamma=(r_1,\dots,r_n)$, of length $|\gamma|=n \in \mathbb{N}^*$, to be a sequence of points in the dual such that $(r_j,\dots,r_n,r_1,\dots,r_{j-1})$ is a path for all $j=1,\dots,n$.  A {\em contour $\gamma$ of the dual of $\Z^2$ is said to occur} in the configuration $\sigma$, or simply to be a {\em contour of} $\sigma$, if it separates clusters of  $+$'s or $-$'s in $\omega$, i.e. if $\gamma \subset \big\{ b + (\frac{1}{2},\frac{1}{2}) : b=\{i,j\}, ||i-j||_1=1, \omega_i \neq \omega_j \big\}$.  To relate Hamiltonians with contours, one observe that the event $\gamma$ that a contour occurs requires an energy proportional to its length $| \gamma |$ ({\em i.e.} the perimeter of the droplet), so that if $\beta$ is large  long contours will be very improbable w.r.t. the probability $\gamma_\Lambda ^{\beta \Phi}(\cdot | +)$. Thus, one relates the energy of a contour  with its length to get
 the following {\em Peierls's estimate:}

% use a physically equivalent lattice-gas potential. Consider  the Hamiltonian with free boundary conditions and rewrite
%\be \label{vacuum}
%H_\Lambda(\omega) = \sum_{\langle ij \rangle \subset \Lambda} \omega_i \omega_j = \sum_{\langle ij \rangle \subset \Lambda} (1-  \omega_i \omega_j) -  \sum_{\langle ij \rangle \subset \Lambda} 1 = \sum_{\langle ij \rangle \subset \Lambda} (1-  \omega_i \omega_j) + C(\Lambda)
%\ee
%where $C(\Lambda)=-\sum_{\langle ij \rangle \subset \Lambda} 1 $. The later is dominated by the volume so that potentials of the form $\omega_i \omega_j$ and $1-\omega_i \omega_j$ are physically equivalent and describe the same phases (see \cite{Geo88, FV16}). 

%\be \label{Peierls}
$$
\gamma_\Lambda^{\beta \Phi}(\gamma| +) \leq e^{-2 \beta |\gamma|}
$$
%\ee

%
From this, thanks to an entropic bound counting the number of contours of a given length, it is possible to estimate  the probability that the spin at the origin takes value $-1$, an event which implies the occurrence of contours, by

\be \label{Peierls2}
\gamma_\Lambda^{\beta \Phi}(\sigma_0=-| +) \leq \sum_{l \geq 1} l 3^l e^{-2 \beta l}
\ee

Using (\ref{mupm}), this yields the weak convergence as $\beta$ goes to infinity of $\mu_\beta^+$  to the Dirac measure $\delta_+$, while the $-$-phase can be similarly proved to converge to the Dirac measure $\delta_-$. 

%One also gets the validity of Peierls's estimate  for the different phases at low enough temperature: the probability that there is a contour which surrounds the origin is bounded by the rhs of (\ref{Peierls2}) while the probability that a given contour $\gamma$ surrounds the origin satisfies
%\be \label{Peierls1}
%\mu_\beta^+(\gamma \ni 0) \leq e^{-c \beta |\gamma|}, \; {\rm for} \; c>0.
%\ee
%This analysis gave rise to the fundamental {\em Pirogov-Sinai theory of phase transitions} for more general models
%\cite{PS,Sin},  under {\em Peierls's conditions} of the form (\ref{Peierls1}). An important property used here is the additivity of Hamiltonians, that could be relaxed a bit as quasi-additivity property used in long-range models (see next section  or \cite{LP2017}).

%Important facts : see $e.g.$ \cite{LP2017, Vu}

%Addivity of the Hamiltonian in terms of contours
%\be \label{additivity}
%H_\Lambda^+ (\underline{\gamma},\gamma_0) = H_\Lambda^+ (\gamma_0) + H_\Lambda^+ (\underline{\gamma})
%\ee
\begin{itemize}
\item{{\bf Phase transition and Dobrushin states in $d=3$}}
\end{itemize}

In our ferromagnetic models, phase transitions at higher dimensions are  implied by those of lower dimensions, by stochastic domination. In particular, for such models, the critical temperature in dimension 3 is at least the one in dimension 2 :  $T_c(d=2)< T_c(d=3)$.

Nevertheless, there could be  intermediate ranges of temperature where the phase diagram could coincide or not with the $2d$-picture : either there are {\em only (2) t.i. extremal Gibbs measures and no non-translation-invariant extremal Gibbs measures}, either there are {\em (at least countably) many's non-t.i. extremal Gibbs measures}\footnote{Dorushin states, got by weak limit with mixed Dobrushin b.c. centered at any plane $\pi$ are known to exist in $3d$, but it is open whether there do exist other non t.i. extremal Gibbs measures.}.

 The $3d$-picture, where there are indeed countably many's non-translation-invariant extremal Gibbs measures, have been first  described by Dobrushin in 1972 \cite{Dob72}. The original idea is to used the mixed so-called {\em $\pm$- Dobrushin b.c.}  (located at the origin), defined such that
\begin{eqnarray*}
\forall x=(x_1,x_2,x_3) \in \Z^3, \; \pm_x &=& +1 \;\; \;  {\rm if} \; \; \; x_1 \geq 0\\
%\\
\pm_x &=& -1 \; \; \; {\rm otherwise}
\end{eqnarray*}
and to prove that the corresponding limiting Gibbs states $\mu^\pm$ cannot be translation-invariant as soon as there is phase transition in $2d$, so for temperatures $T \leq T_c(2)< T_c(3)$. Such a temperature, where some extremal states cease to be translation-invariant, is called the {\em Roughening temperature} (see {\em e.g.} \cite{BLP79, BFL82, FV16}). The infinite-volume limit $\mu^\pm$ would exhibit more coexistence near this plane, and more $+$'s or $-$'s, further up or down from it. This yields a {\em non-translation-invariant  extremal states} $\mu^\pm$, which thus cannot be a convex mixture of the other extremal states $\mu^+$ and $\mu^-$, so in particular
 $$
 \mu^\pm \neq \frac{1}{2} \mu^- + \frac{1}{2} \mu^+
 $$

As a consequence, the microscopic interface separating the $+$'s and $-$'s would not fluctuate much when the volume increases, and stay located near the original plane : one says that this interface is {\em rigid}. 
This construction could be done for any horizontal plane $\pi : x=h$, or even more any plane in $\Z^3$, and thus a countable family of different 'Dobrushin' b.c., so that one gets at least countably many's non-translation-invariant extremal Gibbs measures $\mu_{\pi}^\pm$. In Section 4, we detail a bit more the proof  van Beijeren  provided afterwards  (\cite{vB}, 1976) in the  case of some {\em anisotropic long-range Ising models on} $\Z^2$.

As we shall see now, this rigidity does not hold for $d \leq 2$ for $n.n.$  Ising models \cite{Geo88, Aiz, Hig}, nor for long-range models in $d=1$ \cite{Fannes82, Geo88}, neither for anisotropic long-range Ising models in $d=2$, where  Gibbs measures got by Dobrushin b.c. are not Dobrushin {\em states} : they are either non-extremal, either non-translation-invariant as shown in \cite{CELNR18}, see Section 4.  For more general results on translation-invariant extremal Gibbs measures for finite-range  Ising model, see \cite{Bodineau06, Raoufi18}.

Note that in case of rigidity, this mixed $\mu^\pm$-states give rise to many peculiar measures, such as some local but non global Markov measure \cite{Foll,Gold2,Isr,vW}  or some Gibbs measure which is not the limit of any finite measures with b.c. \cite{Co}.
\subsection{Fluctuations and rigidity of interfaces in the $n.n.$ cases ($d=2,3$)}

	\begin{figure}[ht]
		\centering
		\includegraphics[width=0.33\textwidth]{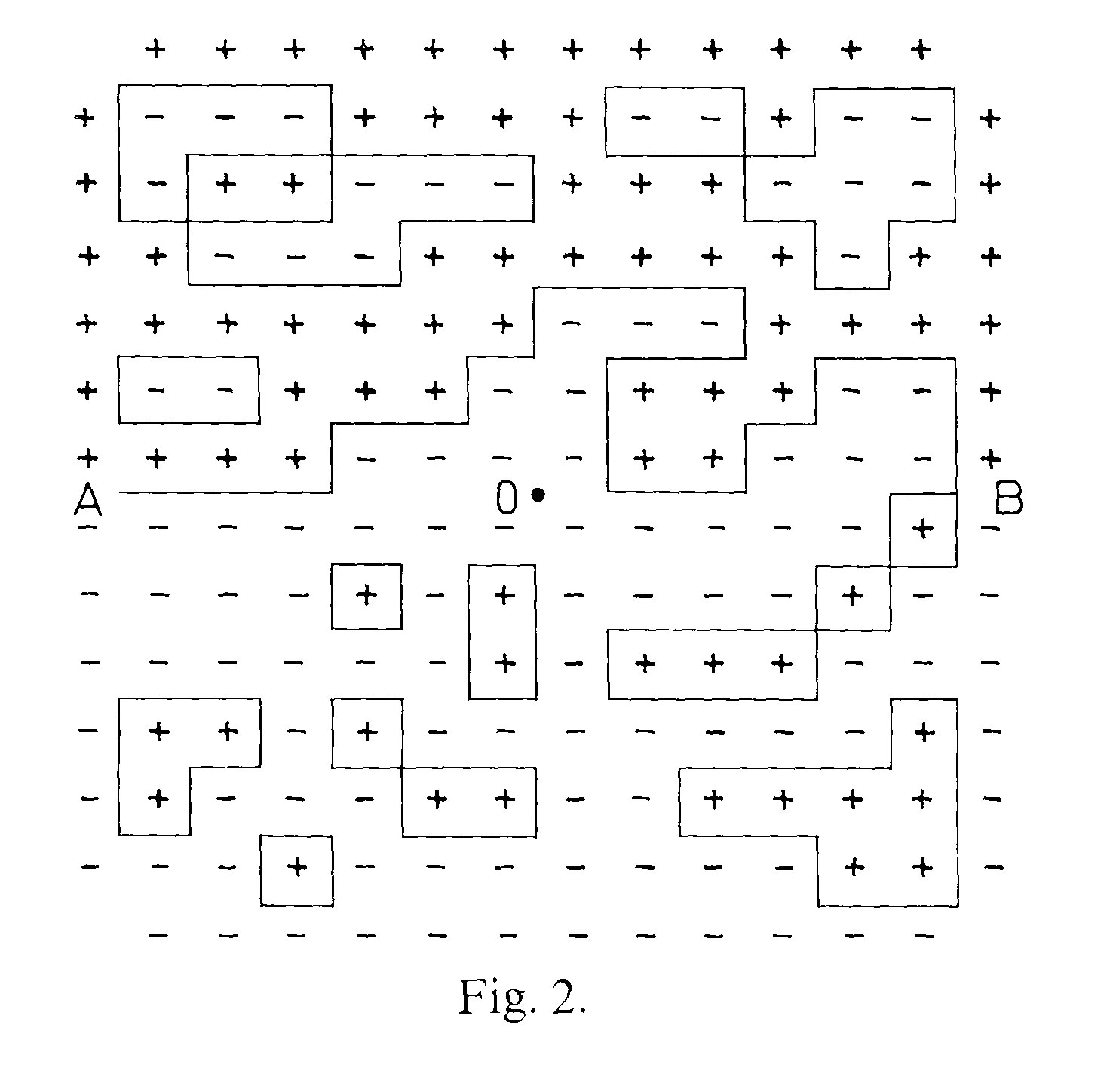}
	\caption{Microscopic interface : Gallavotti line}
		\label{fig-dob}
\end{figure}

\hspace{.5cm} The absence of non-translation-invariant extremal Gibbs measures for the Ising models has been a long-standing case of studies in the seventies. While Dobrushin was formalizing the  $3d$-picture, Gallavotti studied the asymptotic behavior of the microspic interface separating the two phases in the $2d$ case (\cite{Gal72}, 1972). In particular, starting from a square of basis $L$ growing to the whole space $\Z^2$, he proved that with high probability, the interface will  fluctuated at distance $\sqrt{L}$, either up, or either down with equiprobability. This has also been formalized in \cite{MMSole77} who combined these results with correlation inequalities \cite{Leb72,Leb74, Heg77, Leb77} to eventually get a non-extremal but translation-invariant Gibbs measure
\be \label{dob-ti}
\mu^\pm = \frac{1}{2} \mu^- + \frac{1}{2} \mu^+
\ee
On the contrary to dimension 3 where the fluctuations of the interface remain bounded \cite{Dob72}, these fluctuations have been afterwards shown to have a {\em Gaussian profile} by Abraham/Reed \cite{AbrRee1976}, 1976), fluctuating indeed as $\sqrt{L}$ for a box $\Lambda$ of basis $L$. With such boundary conditions, this interface will eventually fluctuate up or down, with probability $\frac{1}{2}$ each, at a ballistic speed. These properties have been extended to many other mixed non translation-invariant boundary conditions one can imagine (See \cite{MMSole77} and references therein). The difficulty afterwards was to be able to prove even for boundary conditions one could not imagine, and even for Gibbs measures that could arise without any boundary condition.

The studies eventually culminate by the works of Aizenman \cite{Aiz} or Higuchi \cite{Hig}, excluding translation-invariant extremal states other than $\{\mu^-,\mu^+\}$ by percolation methods based on a previous work of Russo \cite{Russo79}. This eventually leads to  the full convex picture $\mathcal{G}(\gamma)=[ \mu^-,\mu^+]$ so that the convex decomposition (\ref{Choquet}) reduces to :
$$
\forall \mu \in \mathcal{G}(\gamma) = \alpha_\mu^+ \mu^+ + (1 - \alpha_\mu^+) \mu^-
$$
where the weights $\alpha_\mu^+$ are given by (\ref{weights}).

The behavior of this interface in $2d$ has been refined up to the critical point, see the more precise results by Higuchi \cite{H0,H1}, Greenberg/Ioffe \cite{GI05} or other investigations of Bricmont {\em et al.} \cite{BFL82, BLP81, Pfi91}. See also results got by percolation approach by Gielis/Grimmett \cite{GG}  or more recently by Cerf/Zhou \cite{Cerf}.
\section{Long-range Ising models in dimension one ({\em Dyson models})}
\hspace{.5cm} We briefly describe  the history of these long-range models since its modern introduction in 1969 by Kac and Thompson, where phase transition for decays $1 < \alpha \leq 2$ was conjectured \cite{KT69}. The $(1 < \alpha < 2)$-cases were solved by Dyson at the same time using a bound on the magnetization with this of a hierarchical model \cite{Dys}. He extends its results and partially solved the borderline case  $\alpha=2$  in 1971 \cite{Dys71}, while the complete proof of this rich hybrid case was provided in 1982 by Fr\"ohlich/Spencer \cite{FrSp82}, with afterwards many peculiar properties that we will not describe here, see {\em e.g.} Aizenman {\em at al.} \cite{AN86, ACCN88} or Imbrie {\em et al.} \cite{I82, IN88}. 

\begin{proposition}\label{DyFrSp}
The Dyson  model with specification $\gamma$ and potential (\ref{Dys}) exhibits a phase transition at low temperature  for slow decays $1< \alpha \leq 2$:
$$
\exists \beta_c >0, \; {\rm such \; that} \; \beta > \beta_c \; \Longrightarrow \; \mu^- \neq \mu^+ \; {\rm and} \; \mathcal{G}(\gamma)=[\mu^-,\mu^+]
$$
where the extremal phases $\mu^+$ and $\mu^-$ are translation-invariant. They have in particular opposite magnetisations   $\mu^+[\sigma_{0}]=-\mu^-[\sigma_{0}]=M_0(\beta, \alpha)>0$ at low temperature.
\end{proposition}
It is  known that all Gibbs measures for  Dyson models are translation-invariant  \cite{Geo88,Fannes82}.

Phase transition in these long-range models takes its origin in the possibility, due to the infinite range of the interaction, for the entropy to lose against energy at low temperature, for slow decays $\alpha \leq 2$, thanks to a  
{\bf  volume-dependent energy cost} needed to create a droplet of the opposite phase in a ground-state configuration, as for $n.n.$ Ising models in dimension $d \geq 2$. In these estimates, the dimension $d$ is replaced as a parameter by the decay $\alpha$, so that the latter can be used to tune the dimension, in a continuous manner. See {\em e.g.} \cite{vE19}.

The original estimate was already observed by  Landau/Lifschitz  \cite{LL}, and is sometimes called {\em Landau estimate} \cite{Picco}. 
%It consists, starting from one of the extremal phases, in an estimation of the energetic cost required to insert a droplet of size $2L$ of an opposite phase.
 In our situation, start with the $+$phase, got by monotone weak limit with homogeneous $+$-boundary condition as defined in (\ref{mupm}), for our  pair-potential $\Phi$ long-range couplings $J=J^\alpha$ as in (\ref{Jalpha}), for $d=1$ and $\alpha>1$.

%$$
%\mu^+ (\cdot) = \lim_{\Lambda \uparrow \mathcal{S}} \gamma_\Lambda^{\beta \Phi}(\cdot | +)
%$$
%{\bf Excess energy :} 
Write the excess energy $h_L:=  H_\Lambda(- | +) - H_\Lambda(+ | +)$ at volume $\Lambda=\Lambda_L$  to be the cost of inserting of droplet of the opposite phase, for  finite-volume intervals $\Lambda$ of length $2L$. {\em Landau estimate} tells that the {\em finite-volume  excess  energy} $h_\Lambda$ is has indeed a volume-dependent order:
\begin{equation}\label{LandauEst}
h_L^+ \approx \sum_{j=-L}^L  \sum_{k \geq L}^\infty \frac{1}{k^\alpha} \approx C \cdot L^{2-\alpha}.
\end{equation}

While it had been already been used to get uniqueness  for fast decays $\alpha >2$ \cite{Ruelle68, Thouless69, SS81}, it tells us in particular that the energetic cost to insert  droplet/interval $\Lambda$ of length $L$ of the opposite phase, is volume-dependent for $\alpha \in (1,2)$. Thus, for very long ranges (also called slow decays),  the probability of occurrence of a droplet of the opposite phase is depressed at least by
\begin{equation}\label{CassDepr}
c \exp{-\beta \zeta L^{2-\alpha}}, c,\zeta >0, (\alpha < 2)
\end{equation}
%or 
%\begin{equation}\label{CassDeprBorder}
%c \exp{-\beta \zeta \ln{L}}, c,\zeta >0, (\alpha = 2).
%\end{equation}

The analogy with $d>1$ where the bounds goes as $c \exp{-\beta L^{(d-1)/d}}$ is evident, but we warn the reader that other analogies exist (for {\em e.g.} critical exponents in \cite{AizFer88}).
The results described here can be completed by the concise  introduction of Littin/Picco (\cite{LP2017}, 2017), or any of the introduction in the series of papers of Cassandro {\em et al.} \cite{CFMP,COP09, COP12, CMP17,CMPR}. 

A crucial step to formalize these ideas has been the 2005 paper of Cassandro/Ferrari/\-Merola/Pressuti which provided an explicit and rigorous geometric description of Gibbs measures in the phase transition region. We describe it in next section.

%The bijection between configurations and {\em ad-hoc} contours is written in full generalirt in most of them, and in a very didactic and detailed way in Master thesis of Vu \cite{Vu}.

%uniqueness + G(gamma) : Fannes {\em et al.}  \cite{Fannes82} but also \cite{BS83} using  relative entropy estimate as in \cite{BLP79}(as we do in 2d to deal with the absence of Dobrushin states).
\hspace{.5cm}

%\begin{itemize}
%\item{{\bf G\'eometric Peierls-like proof of phase transition for slow decays}}
%\end{itemize}
\subsection{Triangle-contour construction -- Peierls-like argument}
\hspace{.5cm} In this subsection, we sketch the {\em triangle-contour construction} in  one-dimension, for long-range Ising models with slow decays. Premices of these notions were originally coined to treat the borderline case $\alpha=2$ by Fr\"ohlich/Spencer \cite{FrSp82}, with the introduction of {\em spin-flip} or {\em interface} points, {\em pre-contours} and {\em contours} inspired by the dipole described by the same authors for  the two-dimensional Coulomb gas in \cite{FrSp81}. The geometric description of configurations in terms of these contours lead to a bijection as soon as one leave some possible ambiguities. This was done later by the {\em triangle-contour} description of Cassandro {\em et al.} by randomizing the lengths of the droplets, in order to be able to generate uniquely one geometric construction by configuration. The explicit contruction together to the required quasi-additivity needed to get  Peierls estimate have been afterwards developed in a  series of paper of Cassandro {\em et al.}, starting from  \cite{CFMP} with some technical restrictions\footnote{Large $n.n.$ couplings $J(1)=J >>1$ and a restricted range of decays $1 < \alpha_* < \alpha < 2$ for some $\alpha_* \approx 1,41..$} partially reduced afterwards, either by Littin/Picco \cite{LP2017} or by Bissacot {\em et al.} \cite{BEEKR18}.

A {\em contour} associated with a configuration $\sigma$ will be formed by droplet(s) of the opposite phase, well separated enough so that one recovers some weak subadditivity on their Hamiltonians of the form of (\ref{Sub}). To avoid amibiguities and get a bijection between configurations and contours, the main idea of \cite{CFMP} has been to randomize the length of the droplets, to be able to call them one-by-one in a procedure inspired by coarse-graining in one dimension \cite{Derrida, CarrPego}. We refer to the Cassandro {\em et al.} series of papers and to the thesis of Littin \cite{Littin2013} for the proofs of the bijection configuration-triangle first, and triangle-contours afterwards.

 We first introduce the necessary notions to get such relevant contours in $1d$. Then we describe the Peierls estimate they obtain in this one-dimensional long-range context. In addition, this triangle construction also allows an unambiguous notion of microscopic interface (with mesocopic fluctuations) in the phase transition region, as we shall see in next subsection.

%{\bblue ***
%$J(1) >> 1,\; \alpha>\alpha_+$
%**}
\begin{description}

\item{{\bf Step 1. Bijection configuration-triangles}}
\end{description}
\hspace{.5cm} For $+$-b.c., there is a unique ground state, the $+$ configuration s.t. $+_x=+1,\; \forall x \in \Z$. 

In this one dimensional model, impurities from this ground states are caracterised by the existence of  {\em spin-flip points} $x \in \mathbb{Z}$ on the dual lattice, yielding an interface at $(x,x+1)$ when $\sigma_x \sigma_{x+1}=-1$. Start from a configuration $\sigma$ and enumerate the defects ('$-$') from, say,  the left boundary; the first spin-flip point separates then a row of consecutive plusses to a (maybe singled) row of the opposite phase, which flips again at the next spin-flip point, and so on. One would like to group the rows of defects into classes separated enough to be considered as almost independent, depending on the decay $\alpha$. Triangles are then built on rows of identical spins between two spin-flip points. 

The complete the construction, Cassandro {\em et al.}   provided  an algorithm to get uniquely, from a configuration $\sigma$, a family of triangles
$$
\bar{T} = (T_1,\dots,T_k,\dots,T_n)
$$ 
ordered by lengths 
$$
|T_k| \leq |T_{k-1}|
$$

Triangles $T=T_k$'s are subsets of the dual lattice, whose length $|T|$ is the number of sites imbetween two spin-flip points. The algorithm provided by Cassandro {\em et al.} (see also Picco {\em et al.} \cite{Littin2013,LP2017,Vu}) is such that the triangles of the family $\bar{T}$ satisfy the following properties:
\begin{enumerate}
\item Triangles are well separated one to the other:
$$
dist(T,T') > \min{(|T|,|T'|)}
$$
so that they indeed represent droplets of the opposite phase, with $+$'s imbetween.
\item The associated Hamiltonian $H^+_\Lambda(\bar{T})=H^+_\Lambda(\sigma)$ is additive :
$$
H^+_\Lambda(\bar{T}) = \sum_{k=1}^n H_\Lambda^+(T_k).
$$

\item The energetic cost needed to remove the smaller triangle,
$$
H_k(\bar{T}):=H(T_k,T_{k+1},\dots,T_n) - H(T_{k+1},\dots,T_n)  
$$
satisfies 
\be \label{PeierlsTri}
H_k(\bar{T}) \geq \kappa_\alpha |T|^{2-\alpha}
\ee
with $\kappa_\alpha : = 2(3-2^{3-\alpha})$.
\end{enumerate}

Note that $\kappa_\alpha >0$ only for $1< \alpha_* < \alpha <2$, which is the reason of the original restriction on decays. It had been avoided by Littin/Picco by providing a similar bound for contours (and not triangles, see \cite{LP2017}). For the sake of simplicty, we decribe here the version of the construction with these technical  constraints ($J(1)>>1$ and $\ln{3}/\ln{2}=\alpha_* < \alpha <2$). Note also that in their construction their could be triangles inside triangles.

In such geometric construction, one key points are to avoid ambiguities in the choice of the geometric objects, and second to insure that the process described indeed leads to something. This was done in \cite{CFMP}, pursued and upgrades in the series of papers \cite{COP09, COP12, CMPR, CMP17}, also described in a didactic way in \cite{Littin2013, LP2017, Vu}. 

%The technical conditions crucial in the above works to insure the existence a non trivial bound in (\ref{PeierlsTri}), have been thereafter dropped by Bissacot {\em et al.} or Littin {\em et al.} \cite{Littin2013, LP2017}.

\begin{description}
\item{{\bf Step 2 : "Contours" as bands of nearby "triangles"}}
\end{description}
\hspace{.5cm} The second ingredient needed for Peierls estimate machinery is a subadditvity, of the form:
\be \label{subTri}
H(T_1,T_2) \geq \zeta H(T_1) + H(T_2)
\ee
when $T_1$ and $T_2$ are two different non-overlapping droplets/triangles. As shown in {\em e.g.} \cite{CFMP, Vu}, this cannot be always the case for any pair of triangles. The idea in the definition of contours is the following : if (\ref{subTri}) does not hold, then group the triangles $(T_1,T_2)$  in the same {\em contour}. This is in particular the case when one has

\be \label{sepTri}
dist(T,T') > C(\delta) \min{(|T|,|T'|)}^\delta,\; \delta \geq 1
\ee

so one could group  together the triangles that are too close, in order to form a {\em contour}. The original choice in \cite{CFMP} was made with $\delta =3$, and they indeed describe an algorithm producing a family of contours $\bar{\Gamma}=\bar{\Gamma}(\bar{T})=\bar{\Gamma}(\sigma)$ such that
\begin{enumerate}
\item To a configuration $\sigma$ there corresponds a unique family of contours
$$
\bar{\Gamma}=(\Gamma_1,\dots,\Gamma_2)
$$
where $\Gamma_i=\{ T_{i,j},\; 1 \leq j \leq k_i\}$ is formed by triangles well seprarated from each other, {\em i.e.} statisfying (\ref{sepTri}).

\item The length of the contours is the sums of the lengths of the triangles belonging to it :
$$
|\Gamma| = \sum_{T \in \Gamma} |T|.
$$

\item Contours generated by the triangles generated by a configuration $\sigma$ are themselves also well-separated :
\be \label{sepCont}
dist(\Gamma,\Gamma') > C \min{(\Gamma,\Gamma')}^3
\ee
where 
$$
dist(\Gamma,\Gamma') = \min_{T \in \Gamma, T' \in \Gamma'} dist(T,T')
$$
%\item {\bw Independance of countours....}
\end{enumerate}
Other technical conditions are needed to insure the convergence and uniqueness of the algorithm, see \cite{CFMP}. When two triangles $T$ and $T'$ belonging to different contours have disjoint support, one says that they are mutually external, but this is not always the case, see \cite{CMPR,LP2017}.

\begin{description}
\item{{\bf Step 3 : Quasi-additive bounds of the Hamiltonians}}
\end{description}
%-- generation of triangles (droplets fact), consequences on the distance etc.
\hspace{.5cm} To get its estimate, and avoid too strong dependencies between contours, Peierls used
$$
H(\Gamma_0,\Gamma_1,\dots,\Gamma_n) = H(\Gamma_0) + H(\Gamma_1,\dots,\Gamma_n) 
$$
but in fact the following weak form of  subadditivity is enough :
\be \label{Sub}
H(\Gamma_0,\Gamma_1,\dots,\Gamma_n) \geq \zeta H(\Gamma_0) + H(\Gamma_1,\dots,\Gamma_n), 0< \zeta < 1.
\ee
This was proved in \cite{CFMP} for slow decays and extended to the whole range of decays $1 < \alpha \leq 2$ by Littin {\em et al.}, with an extension of Landau estimate to contours, with a removing cost estimated as:
\be \label{LandTri}
H(\Gamma) \geq \zeta_\alpha \sum_{\bar{T} \in \Gamma} |T|^{2-\alpha}
\ee
with $\zeta_\alpha>0$ for $\alpha_*<\alpha< 2$  (for other decays, a mixed energy-entropy argument is needed \cite{CFMP,LP2017,BEEKR18}).

%********From COP09***
\begin{description}
\item{{\bf Step 4 : Peierls argument}}
\end{description}
\hspace{.5cm} A necessary condition to have $\sigma_0=-1$ is that the origin $0$ is contained in the support of some contour $\Gamma$, so that :
$$
\mu^+_\Lambda \big(\sigma_0 =-1 \big) \leq \mu_\Lambda^+ \big( \{ \exists \Gamma : 0 \in \Gamma \}\big) \leq \sum_{\Gamma \ni 0} \mu_\Lambda (\Gamma)
$$
and, using (\ref{LandTri}),  relate it to the lengths of the triangles to get:
$$
\mu^+_\Lambda \big(\sigma_0 =-1 \big) \leq \sum_m \sum_{\Gamma : |\Gamma| = m, 0 \in \Gamma} e^{-\frac{\beta \zeta}{2} \sum_{T \in \Gamma} |T|^{2-\alpha}}
$$
To conclude, on uses an entropy estimate counting the number of such triangles \cite{CFMP} to get  for $m\geq 1$ and some $b$ large enough, 
$$
\sum_{\Gamma : |\Gamma| = m, 0 \in \Gamma} e^{-\frac{\beta \zeta}{2} \sum_{T \in \Gamma} |T|^{2-\alpha}}  \leq 2 m e^{-bm^{2-\alpha}}
$$
and eventually phase transition for $\beta$ large enough.

%\end{description}

%In fact, there are some further applications to inhomogeneous situations where this the large nearest-neighbour condition alreaDynkin78 was removed, see \cite{BEvEL}.

%The restriction of $\alpha>\alpha_+$ appears since in \cite{CFMP} the proof of the phase transition of the Dyson model by a contour argument needs it\footnote{Although for the existence of a transition the validity can be extended to the whole range of phase-transition decays by FKG arguments. This does not work for  inhomogeneous situations such as disordered systems \cite{COP09} or interface fluctuations \cite{CMPR}.}, while the contours introduced are based on the triangles  defined above.

\subsection{Non-Gibbsianness in $1d$ : decimation of Dyson models}\label{SectionNG} 

\hspace{.5cm} A particular consequence of this phase transition is that it  provides an example of a non-Gibbsian measure in dimension one, briefly described here (see also \cite{ELN17}).

In this subsection, we use the well-known characterization of Gibbs measures as being {\em quasilocal} and {\em non-null}. Quasilocality is a Feller-type property equivalent to the existence of continuous versions of conditional probabilities, in the product topology of the discrete one on $E$, providing  an interpretation of  Gibbs measures as natural extensions  Markov fields.  The rigorous proof of the equivalence  was coined by Kozlov (\cite{Koz74}, 1974) and Sullivan (\cite{Sull76}, 1976). Note that in one implication (from quasilocality to Gibbsianness), some non-trivial issues about translation-invariance arise, see discussion in {\em e.g.}  \cite{EFS93,Fer,LN09}

%Gibbs measures : quasilocal and non null}

When $\mu \in \mathcal{G}(\gamma)$ is quasilocal, then for any  $f$ local and  $\Lambda \in \s$,  the conditional  expectations of $f$ w.r.t. the outside of $\Lambda$ are $\mu$-a.s. given by $\gamma_\Lambda f$, by  (\ref{DLR}), and this is itself a continuous function of the boundary condition by (\ref{GibbsSpe}). Thus, one gets for any $\omega$
\be \label{esscont}
\lim_{\Delta \uparrow \mathbb{Z}} \sup_{\omega^1,\omega^2 \in \Omega}  \Big| \mu \big[f |\mathcal{F}_{\Lambda^c} \big](\omega_\Delta \omega^1_{\Delta^c}) - \mu \big[f |\mathcal{F}_{\Lambda^c} \big](\omega_\Delta\omega^2_{\Delta^c})\Big|=0
\ee
%which yields an (almost-sure) asymptotically  weak dependence on the conditioning, which can be seen as an extended  Markov property. 
%In particular, for Gibbs measures the conditional probabilities always have continuous versions, or equivalently
%In particular, for Gibbs measures, it is not possible to change  its conditional probabilities on exceptional sets in order to get a discontinuous version: one says that 
% there is no point of essential discontinuity, in the   sense that it cannot be changed on a positive neighborhood to a continuity point, see \cite{ELN17}. 
 
%One also says that such a configuration is a {\em bad configuration} \cite{EFS93,LN09}.
 
%The existence of such bad configurations implies non-Gibbsianness of the associated measures.

As described in whole generality by van Enter {\em at al.} \cite{EFS93}, this does not always hold for {\em renormalized}  Gibbs measures; Let us describe now the simple such transformation  leading to essential discontinuity when applied to, so-called {\em decimation}.

\smallskip
{\bf Decimation Transformation:} It is defined on the configuration space as
\smallskip
\be \label{DefDec}
 T \colon (\Omega,\mathcal{F})  \longrightarrow (\Omega',\mathcal{F}')=(\Omega,\mathcal{F}); \; 
\omega \; \;   \longmapsto \omega'=(\omega'_i)_{i \in
\mathbb{Z}}, \; {\rm with} \;  \omega'_{i}=\omega_{2i}
\ee
It acts on measures in a canonical way: denote $\nu^+:=T \mu^+$ the decimation of the $+$-phase
% It is formally defined as an image measure via
$$
\forall A' \in \mathcal{F'},\; \nu^+(A')=\mu^+(T^{-1} A')=\mu^+(A) \; {\rm where} \; A=T^{-1} A'= \big\{\omega: \omega'=T (\omega) \in A' \big\}.
$$
%When necessary, we distinguish between original and image sets using  primed notation\footnote{Notice that by  rescaling  the configuration spaces $\Omega$ (original) and $\Omega'$ (image) are identical.}.

%This type of transformation was also the basic example in 
In the seminal work of van Enter/Fern\'andez/Sokal (\cite{EFS93}, 1993),  non-quasilocality of the decimated measure $\nu^+$ is proved in dimension 2 at low enough temperature, as soon as a phase transition is possible for an Ising model  on the {\em decorated lattice}, which consists of a version of $\mathbb{Z}^2$ where the "even" sites have been removed. Here, the role of the image 'decorated' lattice    is 
%equivalent {\em to} %of 
played by the set of odd sites, $2 \mathbb{Z}+1$, which  can be identified with 
$\mathbb{Z}$ itself, and  when a  phase transition holds for the Dyson specification -- thus at low enough temperature for $1 < \alpha \leq 2$ -- the same is true for a constrained specification  {\em with alternating constraint} due to the alternating configuration, yielding non-Gibbsianness of $\nu^+$.

% Once the $+$-phase is shown to be non-Gibbsian after being subjected to a decimation transformation, the same holds true for all other Gibbs measures of the model.  
% We shall come back to this later, before we state and prove our main result. 

%{\bf Non-Gibbsianness at Low Temperature}
\begin{theorem}\label{thm2}\cite{ELN17}
For any  $1<\alpha \leq 2$, at low enough temperature,
% $\beta > \beta_c^D$, 
the decimation  $\nu=T \mu$ of any 
%$+$-phase
 Gibbs measure $\mu$ of the Dyson model is non-quasilocal, hence non-Gibbs.
\end{theorem}

For the full proof, see \cite{ELN17}. Here, we only pick-up a sketch of the proof.

The point of essential discontinuity we exhibit, called the {\em bad configuration} for the image measure $\nu^+$ is  the  {\em alternating configuration} $\omega'_{{\rm alt}}$ defined for any $i \in \mathbb{Z}$ as $(\omega'_{\rm alt})_i=(-1)^i$. To get the essential discontinuity, the choice of $f(\sigma')=\sigma'_0$ and conditioning outside $\{0\}$ will be enough. Due to cancelations and symmetries, conditioning by this alternating configuration yields a constrained model that is again a model of Dyson-type which has a low-temperature transition in our range of decays $1 < \alpha \leq 2$. The  proof essentially goes along the lines sketched in \cite{EFS93,ALN}, with the role the ``annulus'' played by two large intervals $[-N,-L-1]$ and $[L+1,N]$ to the left and to the right of the central interval $[-L,+L]$. If we constrain the spins in these two intervals to be either plus or minus, within these two intervals the measures on the unfixed spins are close to those of the Dyson-type model in a positive, or negative, magnetic field. As those measures are unique (\cite{YL, Kerimov2009}) no influence from the boundary can be transmitted by via the ``annulus''.. However, due to the long range of the Dyson interaction, there may be also a direct influence from the boundary to the central interval. To overcome this difficulty, we choose $N(L)$ large enough as $N = L^{\frac{1}{\alpha -1}}$, in order to  make this direct  influence as small as he wants.

The main tool to justify this rigorously is to consider the ''Equivalence of boundary conditions'' concept  coinded by Bricmont/Lebowitz/Pfister in the beautiful paper \cite{BLP79}, by considering b.c. $\omega'^\pm$ either in the $+$- or in the $-$-neighbourhoods of the alternated configuration. Write $\Lambda'=\Lambda'(L)=[-L,+L]$ and $\Delta'=\Delta'(N)=[-N,+N]$, with $N >L$ and denote formally  by $H$ the Hamiltonian of the constrained specifications for $\omega_1^+$ and $\omega_2^+$ as prescribed. One can bound uniformly in $L$ the relative Hamiltonians as 
\be\label{bc}
\Big| H_{\Lambda,\omega_1^+}(\sigma_\Lambda) - H_{\Lambda,\omega_2^+}(\sigma_\Lambda) \Big| \leq C <  \infty.
\ee
as soon as one takes $N=N(L)=O(L^{\frac{1}{\alpha -1}})$. Then one gets by \cite{BLP79} (see also \cite{FV16}) that all of the limiting Gibbs states obtained by these boundary conditions have the same measure zero sets, an equivalent decomposition into extremal Gibbs states (presumably trivial here, as the Gibbs measure will be unique, as we shall see), and thus yield the same magnetisation : $M^+=M^+(\omega, N, L)= M^+(\omega_1^+, N, L)=M^+(\omega_2^+, N, L)$ is indeed independent of $\omega$ as soon as it belongs to the pre-image of the $+$-neighboorhood of the alternating configuration. To get (\ref{bc}), we use the long-range structure of the interaction to get a uniform bound
$$
\Big| H_{\Lambda,\omega_1^+}(\sigma_\Lambda) - H_{\Lambda,\omega_2^+}(\sigma_\Lambda) \Big| \leq 2 \sum_{x=-L}^L \sum_{k > N} \frac{1}{k^\alpha} < 2 L \frac{N^{1-\alpha}}{1-\alpha}
$$
so that choosing $N=N(L)$ such that $2 L \frac{N^{1-\alpha}}{\alpha - 1}=1$  will yields the seeked essential discontinuity, so one can choose
\be \label{N}
N(L)=L^{\frac{1}{\alpha-1}}.
\ee
%For example, for $\alpha=\frac{3}{2}$, one has thus to take some annulus of the order of $N(L)=O(L^2)$.

Once we got rid of any possible direct asymptotic effects due to the long range by choosing a large enough annulus as above, the main point is now that 
%the frozen condition can 
freezing the primed spins to be minus can overcome 
%beat the weak limit with
the $+$-boundary condition when the frozen annulus $\Delta' \setminus \Lambda'$ is in a $-$-state, for $L$ and $N(L)$ large enough. The corresponding  magnetization can then be made as close as possible to the magnetisation of the Dyson  model with an homogeneous external field $h_x=-$ everywhere, which at low enough temperature  is smaller than and close to the magnetisation of the Dyson  model under the $-$-phase, i.e to $-M_0(\beta,\alpha) <0$ (and this $-$-phase is also unique). The magnetisation with the constraint $\omega^+$ will thus  be close to or bigger than $+M_0(\beta,\alpha)$ so that a  non-zero difference is created at low enough temperature.

Note that this non-Gibbsianness might be of some importance in the use of renormalization group in  Neurosciences simulations, see \cite{CLNL17} and references therein.

\subsection{Mesoscopic interfaces and (non-) $g$-measure property}

\hspace{.5cm} An another important consequence of the arising phase transition in one-dimension for long-range model with slow decays is the ocurrence of mesoscopic fluctuations of the interface (point) got with mixed $_+$-Dobrushin b.c. ($-$ on the left side of the integrer line, $+$ on the other side). As we show in \cite{BEELN18}, these fluctuations implies a wetting phenomena (propagation of a droplet of the opposite phase), which have itself an important consequence on the continuity properties of one-sided conditional probabilities, providing a seemingly first example of Gibbs measure which is not a $g$-measures in \cite{BEELN18}. We describe this result here; it requires to describe the interface fluctuations results of \cite{CMPR}, and the intermediate wetting consequences also derived in \cite{BEELN18}.

%\smallskip
\subsubsection*{\bf Dobrushin boundary conditions and {\em Interface point}:}
%\smallskip

\hspace{.5cm} For homogeneous boundary conditions, since the number of spin-flip points is even, every spin-flip point was  an extremity of some droplet/triangle. If we consider now a Dobrushin-type boundary condition, then the number of spin-flip points becomes odd, and so there exists a {\em unique} spin-flip point which is not the vertex of any triangle. 

This point is called the {\em interface point}. To describe where it can be located, let discretise the interval $[-1,+1]$ as
$$
T_L=\left\{ -1-\frac{1}{2L},-1+\frac{1}{2L},\ldots,-\frac{1}{2L},\frac{1}{2L},\ldots,1+\frac{1}{2L} \right\},
$$
and consider the  '$-+$'-Dobrushin boundary condition. Given a configuration $\omega$ in $\Lambda=\Lambda_L$, let  $I^*\equiv I^*(\omega)\in \Lambda^*$ be the interface point of the configuration $\omega$, and for $\theta\in T_L$, denote by
$
\mathcal{S}_{\Lambda,\theta}=\{\omega: I^*=\theta L\}
$
the set of configurations in $\Lambda$ for which the interface point lies at  $\theta L$.

Define now  for each $\theta \in T_L$ the probability  to have an interface in $\theta L$ by
$$
\mu^{-+}_{\Lambda}[I^*=\theta L]=\frac{Z^{-+}_{\theta,\Lambda}}{Z^{-+}_{\Lambda}},
$$
where the partition functions $
Z^{-+}_{\theta,\Lambda}=\sum_{\omega \in \mathcal{S}_{\Lambda,\theta}}e^{-\beta H_\Lambda^{-+}(\omega)}\; {\rm and} \;Z^{-+}_{\Lambda}=\sum_{\theta\in T_L}Z^{-+}_{\theta,\Lambda}
$ are defined via the Hamiltonian $H_\Lambda^{-+}$ in volume $\Lambda$ with Dobrushin boundary conditions. For $i\in\Lambda$, the conditional expectation of $\omega_i$, given $I^*=\theta L$, is then 
$$
\mu^{-+}_{\theta,\Lambda}[\omega_i]:=\mu^{-+}_{\Lambda}[\omega_i|I^*=\theta L]=\frac{1}{Z^{-+}_{\theta,\Lambda}}\sum_{\omega \in \mathcal{S}_{\Lambda,\theta}}\omega_i e^{-\beta H_\Lambda^{-+}(\omega)}.
$$
The expectation of $\omega_i$ can then be written in terms of  $\mu^{-+}_{\theta,\Lambda}[\omega_i]$ as
\begin{equation}\label{chain}
\mu^{-+}_{\Lambda_L}[\omega_i]=\sum_{\theta\in T_L}\mu^{-+}_{\theta,\Lambda_L}[\omega_i]\mu^{-+}_{\Lambda_L}(I^*=\theta L).
\end{equation}

%These  constructions of triangles and associated contours are also used in \cite{CMPR} to  get cluster expansions of partition functions that yield first the following proposition, which will be an essential tool for us. Let $Z^{-}_{\Lambda}$ be the partition function on $\Lambda$ with minus boundary condition, and let $\zeta(\alpha)=\sum_{k=1}^{\infty}\frac{1}{k^{\alpha}}$ be the Riemann zeta function. 

Most of the results of this section are based on a convergent cluster expansion for partition functions from \cite{CFMP, CMPR}, where one in particular learns: 

\begin{proposition}\label{prop:CMPR} 
For all $\alpha\in (\alpha_*,2)$, there exists $\beta_0(\alpha) { >0}$ s.t. for all $\beta>\beta_0$ and $\theta \in T_L$,
$$
\log Z^{-+}_{\theta,\Lambda} - \log Z^{-}_{\Lambda}
= -c_L(\alpha)L^{2-\alpha}+e^{-2\beta(\zeta(\alpha)+J)}\frac{L^{2-\alpha}}{(2-\alpha)(\alpha-1)}f_{\alpha}(\theta)(1\pm e^{-c_1(\alpha)\beta})(1+o(L))
$$
where $\zeta(\alpha)=\sum_{k=1}^{\infty}\frac{1}{k^{\alpha}}$ is the Riemann zeta function and $f_{\alpha}(\theta)=(1+\theta)^{2-\alpha}+(1-\theta)^{2-\alpha}$, $c_L=c_L(\alpha) >0$, $c_1=c_1(\alpha)$,  and $J=J(1) \gg 1$.
\end{proposition}
The estimation of  expectation under the $+$-phase has also been estimated in  \cite{CMPR}:
\begin{theorem}
For all $\alpha\in (\alpha_*,2)$, $\exists \beta_0(\alpha)$, $c_1>0$ s.t. $\forall \beta\ge \beta_0$, uniformly in  $\Lambda\Subset \mathbb{Z}$,
\be
\mu^{+}_{\Lambda}[\omega_i]=1-\left[ 2e^{-2\beta (\zeta(\alpha) +J)}\left( 1\pm e^{-c_1(\alpha)\beta } \right) \left(1+o\left(1\right)\right)\right], \;  {\rm for \; all} \;  i\in \Lambda
\ee
\end{theorem}
Thus, after taking the infinite-volume limit,  at low temperatures, the magnetisation satisfies:
\be\label{magnetization}
1-\left[ 2e^{-2\beta (\zeta(\alpha) +J)}\left( 1+ e^{-c_1(\alpha)\beta} \right)\right]\le \mu^{+}[\omega_i] \le 1-\left[ 2e^{-2\beta (\zeta(\alpha) +J)}\left( 1- e^{-c_1(\alpha)\beta} \right)\right].
\ee

%From these rigorous estimates, one can conclude from \cite{CMPR} 
 %one deduces in Corollary \ref{coro:CMPR} below 
 %that the interface point is located in the middle of the interval of $\Lambda$, up to a Gaussian correction which grows sublinearly in $L$. This means that the correction describes mesoscopic fluctuations. In particular, this implies that macroscopic fluctuations are extremely improbable.
%The constraint $J=J(1)\gg 1$ should be superfluous in our paper and in all subsequent papers after \cite{CFMP}.  In fact, there are some further applications to inhomogeneous situations where this the large nearest-neighbour condition already was removed, see \cite{BEEKR18}.
%\smallskip
\subsubsection*{\bf Consequence of interface fluctuations : Wetting transition} 
%\smallskip

\hspace{.5cm} For a fixed $N> 1$,  consider the $+$-phase $\mu^{+}$, conditioned on the event $-_{-N,-1}$ of the occurence of a droplet of $-$'s in an interval $[-N,-1]$. Then we claim in  \cite{BEELN18} that there are two intervals of length of order  $L$, left and right of the fixed interval and of the form $\big[-N-\frac{(1-sL^{\frac{\alpha}{2}-1})}{2}L,-N-1 \big]$, and $[0,\frac{(1-sL^{\frac{\alpha}{2}-1})}{2}L]$, such that for $N \gg L$ both large enough, satisfying $LN^{1-\alpha}=o(1)$, the magnetisation of the spins in one of these intervals conditioned on the event $\{ \omega_{-N,-1}=-_{-N,-1}\}$ is negative. These intervals play the role of a ``completely wet region'' in a wetting transition.
%(See Figure \ref{fig:wetting})
%In other words,

\begin{proposition}\label{prop}
Let $\alpha\in (\alpha_*,2)$ and $\beta_0\equiv \beta_0(\alpha)$ as above. Then, there exists $\beta_1>\beta_0$ such that, for any $\beta>\beta_1$, there exist $s=s(\beta,\alpha)$, $\lambda=\lambda(\beta,\alpha,s)>0$ and $L_0\equiv L_0(\alpha,\beta)\ge 1$ such that, for any $L>L_0$, there exists $N_0(L)>L$ such that, for any $N\ge N_0(L)$, 
\be
\mu^{+} (\omega_i| -_{-N,-1}) \leq -\lambda m,
\ee
for every $i\in [-N-\frac{(1-sL^{\frac{\alpha}{2}-1})}{2}L,-N-1]\cup [0,\frac{(1-sL^{\frac{\alpha}{2}-1})}{2}L]$,
where $m=\langle \omega_0 \rangle^{+}>0$.
\end{proposition}

%\begin{proof}

%Fix $\alpha\in (\alpha_+,2)$ and $\beta_0\equiv \beta_0(\alpha)$ from Proposition \ref{prop:CMPR}. We will first prove the statement for $i\in [0,\frac{(1-sL^{\frac{\alpha}{2}-1})}{2}L]$.

The main idea of our proof is to choose $N$ large enough for the total influence of all spins left of the interval to be bounded by a (small) constant, so that one can neglect boundary effects beyond $-N$ by equivalence of boundary conditions as in \cite{BLP79,ELN17}. Then inside the interval of length $L$, the interface separating the $+$- and $-$ phases is  w.h.p. within the same window as with the Dobrushin boundary conditions. If afterwards we move the $+$-boundary to the right, the location of the interface, by monotonicity, can also move only to the right, that is away from the frozen interface.

\subsubsection*{\bf Consequence of wetting : discontinuity of 1-sided conditional probabilities }
%\medskip
%(Non-) $g$-measure}
%\medskip
%This part comes from \cite{BEELN18} (....). 

%\Large{

\hspace{.5cm} We deduce from the wetting transition the discontinuity of any $g$-function associated with $\mu^+$, which in turn cannot be a $g$-measure. Let us first introduce a bit more $g$-functions and $g$-measures in our context.

In Dynamical systems, similarly to Gibbs mesures in mathematical statistical mechanics, $g$-measures are defined by  combining  topological and measurable notions, with the introduction of  transition functions (the `$g$'-functions) having to be continuous functions of the {\em past only}. One  requires continuity of single-site {\em one-sided} conditional probabilities and says that $\mu$ is a $g$-measure if there exists a (past-measurable) {\em continuous}  and non-null  function $g_0$ which gives ``one-sided'' conditional probabilities, that is non-null conditional probabilities for events localised on the right half line (the ``future"), given a boundary condition  fixed only to the left (the ``past"). To formalize it, define $T:\{-1,+1\}^{(-\infty,0]} \to \{-1,+1\}^{(-\infty,0]}$ be the {shift} $(Tx)_n = x_{n-1}$. Denote by $\mathcal{P}$ the class of positive \emph{$g$-functions} $g:\{-1,+1\}^{(-\infty,0]} \to (0,1]$ such that $\sum_{y\in T^{-1}x}g(y)=1$,  for all $x\in \{-1,+1\}^{(-\infty,0)}$.
%These functions are called 
% Since $\{-1,+1\}^{(-\infty,0]}$ is metrizable, we can define \emph{continuity} of a $g$-function. 
%A $g$-function is called \emph{strongly non-null} if
%\be
%\inf_{x\in \{-1,+1\}^{(-\infty,0)}}g(x)>0.
%\ee 
% }
We shall use  the \emph{past} and \emph{future} $\sigma$-algebras $\mathcal{F}_{<0}$ and $\mathcal{F}_{>0}$  generated by the projections indexed by negative and positive integers. 
\begin{definition}
A probability measure is a $g$-measure, if there is a non null {\em continuous} $g$-function $g_0$, defined on the left (``past'')  half-line configuration space, such that, for each $\omega_0\in \{-1,+1\}$ and $\mu$ a.e. $\tau=(\tau_j)_{j<0}\in \{-1,+1\}^{(-\infty,0)}$,
\be \label{g}
 \mu[\omega_0|\mathcal{F}_{<0}](\tau):= \E_\mu \big[\mathbf{1}_{\sigma_0=\omega_0} | \mathcal{F}_{<0} \big](\tau) = g_0(\tau \omega_0).
\ee
\end{definition}
%In this situation, the function $g:=g_0$ is called a $g$-function.
 For translation-invariant measures, it is extended to any site $i$ with conditional probabilities w.r.t. to the past at site $i$ given by $g_i=g$.

Discontinuity of any candidate $g^+$ to represent a $g$-function for $\mu^+$ will be a consequence of the  entropic repulsion phenomenon describe above. In the following lemma from \cite{BEELN18}, $\mu^{+,\omega}_{\mathbb{Z}_+}[\cdot]$ denotes expectations under a measure $\mu^{+,\omega}_{\mathbb{Z}_+}$ constrained to be $\omega$ on $\Z^-$, with $+$-b.c. otherwise. The neighborhoods $\mathcal{N}_{N,L}^{+,{\rm left}}(\omega_{\rm alt})$ (resp. $\mathcal{N}_{N,L}^{-,{\rm left}}(\omega_{\rm alt}) $) are the configurations which coincide with the alternate configuration with $+$-b.c. (resp. $-$b.c.) beyond $N>L$.

 \begin{lemma}\label{keylemma2}
Consider the  alternating configuration $\omega_{\rm alt}=\big((\omega_{\rm alt})_i\big)_{i\in \mathbb{Z}}$ defined by $(\omega_{\rm alt})_i=(-1)^i$, and take a Dyson model with polynomial decay $\alpha_* <\alpha<2$ at sufficiently low temperature. 
%Let $L<N$ and consider two arbitrary  configurations $\omega^+ \in \mathcal{N}_{N,L}^{+,left}(\omega_{\rm alt})$ and $\omega^{-} \in \mathcal{N}_{N,L}^{-,left}(\omega_{\rm alt}) $. 
Then, there exist $L_0\ge 1$ and $\delta >0$ such that for any  $L>L_0$ there is an $N > L$, with $LN^{1-\alpha}=o(1)$, such that for every  $\omega^+ \in \mathcal{N}_{N,L}^{+,{\rm left}}(\omega_{\rm alt})$ and $\omega^{-} \in \mathcal{N}_{N,L}^{-,{\rm left}}(\omega_{\rm alt}) $,
\be \label{keymagn2}
\left| \mu^{+,\omega^+}_{\mathbb{Z}_+}[\sigma_0] -  \mu^{+,\omega^-}_{\mathbb{Z}_+}[\sigma_0] \right| > \delta.
\ee
\end{lemma}

As a corollary, we obtain the main result of \cite{BEELN18}:

\begin{theorem}\label{thm:main}
 For $\mu$ being either the $+$- or the $-$-phase of  a Dyson model with decay $\alpha_* <\alpha<2$ at sufficiently low temperature, the one-sided conditional probability  $\mu[\omega_0|\mathcal{F}_{<0}](\cdot)$ is essentially discontinuous at $\omega_{\rm alt}$. Therefore, none of the Gibbs measures $\mu$  for the Dyson model in this phase transition region is a $g$-measure.
\end{theorem}

To describe the $g$-functions, we need regular versions of conditional probabilities given the outside of infinite sets, because so is the past (it is the complement of $\Z^-$, whose conditional probabilities are not provided by the DLR equations). Various constructions of such {\em Global specifications} \cite{FP97, ALN, ELN17, BEELN18} to represent these regular versions eventually allow us to consider, for given pasts, the expression of the  $g$-functions  as the magnetisations of Dyson models under various conditionings, see Equation (\ref{gplus}) below. Studying continuity reduces in fact to studying the stability of  interfaces when changing the boundary conditions arbitrary far away in the past.

Starting from $\mu^+$, we introduce $g^+$ to be the candidate to be the $g$-function representing (a version of) the single-site conditional probabilities  (\ref{g}) as a function of the past. Just as in  \cite{FP97,ELN17}, we introduce thus for any ``past'' configuration $\omega \in \Omega$:
$$
g^+(\omega):=\mu^+\left[\omega_0 | \mathcal{F}_{<0}](\omega)\right.
$$
Using the expression  in terms of global specifications (see \cite{FP97, ALN, ELN17}) and constrained measures, one gets, $\mu^+$-a.s. ($\omega$), the following candidate:

\begin{equation}\label{gplus}
g^+(\omega) = \mu_{\Z^+}^{+,\omega} \otimes \delta_{\omega_{({\Z^+})^c}}[\omega_0]
\end{equation}
where $\mu_S^{+,\omega}$ is the constrained measure on $(\Omega_S,\mathcal{F}_S)$ for $S=\Z^+$ here.
%as the (well-defined) weak limit
%\be \label{constrLimit0}
%\mu_S^{+,\omega}(d \sigma_S):=\lim_{\Delta \uparrow S} \gamma^D_\Delta (d \sigma\mid +_S \omega_{S^c}).
%\ee

Previous works, using monotony and right-continuity \cite{FP97, ELN17}, insure that $\mu^+$ is then indeed ``specified'' by $g^+$, in the sense that it is invariant by its left  action: $\mu^+ g^+=\mu^+$.

%{\bf Note:} 
%A non-continuous (= non-regular) $g$-function gives rise to a measure which is NOT a $g$-measure. To be a ``proper'' $g$-function of the past, we would need that in addition to consistency, the function $g^+$ is {\em regular}, i.e.  essentially continuous (for which all  possible discontinuity points can be removed by modifications on  negligible sets).

To prove that $\mu^+$ is not a $g$-measures, we prove that  $g^+$  can take significantly different values on  sub-neighborhoods $\mathcal{N}_{N,L}^{\pm,  {\rm left}}(\omega_{{\rm alt}}) \subset \mathcal{N}_{ L}(\omega_{{\rm alt}})$, for $L$ large and $N$ larger. To do so, we
introduce the  particular  alternating  configuration $\omega_{{\rm alt}}$. To prove that it is a bad configuration, one should find two sub-neighborhoods on which  the value of $g^+$  differs. 

%We  consider first finite-volume approximations of the constrained measure $\mu^{+,\omega}_{\mathbb{Z}^+}$ built as the weak limit with $+$-boundary condition by taking intervals $I_n$ arbitrarily large, larger than any other finite volumes encountered in this paper.

Consider the sub-neighborhoods $\mathcal{N}_{N,L}^{{ \pm},{\rm left}}(\omega_{{\rm alt}})$ for $L<N$, whose size is adjusted later. All together, this leads us to consider a partially frozen Dyson model, either frozen into $+$ outside $I_n$, or %, either into some arbitrary $\omega$ in $[-n,-N]$, or  
into $-$  in the ``annulus'' $[-N, -L]$, and the alternating one $\omega_{{\rm alt}}$ in $[-L,-1]$.

\begin{center}
 \begin{tikzpicture}[]
    
\draw (-6,0) -- (6,0);

\node [below] at (-1.5,-0.2) {$-L$};

\node [below] at (-4,-0.2) {$-N$};

\node [below] at (5.5,-0.2) {$$};

%\node [below] at (-5.5,-0.2) {$-n$};

%------------- Between 0 and L
\foreach \n in {-3,...,3}{%
        \draw[fill] (\n,0) circle (1pt); 
    }
    
\foreach \n in {-6,...,6}{%
        \draw[fill] (\n/2,0) circle (1pt);    
    }
%--------------

%-------------- Between -N and 0
%\foreach \n in {3,...,4.5}{%
%        \draw[fill] (-\n/2,0) circle (1pt) node [above] {$\omega$};        
        
\foreach \n in {6,...,8}{%
        \draw[fill] (-\n/2,0) circle (1pt) node [above] {$-$};    
    }
    
    \foreach \n in {9,...,10}{%
            \draw[fill] (-\n/2,0) circle (1pt) node [above] {$+$};   
    %    \draw[fill] (-\n/2,0) circle (1pt) node [above] {$\omega$};    
    }
%--------------

%-------------- Left -N
%\foreach \n in {1}{%
 %       \draw[fill] (-\n,0) circle (1pt) node [above] {$+$}; 
  %  }
    \foreach \n in {6}{%
        \draw[fill] (-\n,0) circle (1pt) node [above] {$+$}; 
    }
\foreach \n in {11}{%
        \draw[fill] (-\n/2,0) circle (1pt) node [above] {$+$};    
    }
    
       \foreach \n in {6}{%
        \draw[fill] (\n,0) circle (1pt) node [above] {$+$}; 
    }
\foreach \n in {11}{%
        \draw[fill] (\n/2,0) circle (1pt) node [above] {$+$};    
    }
    
       \foreach \n in {6}{%
        \draw[fill] (\n,-2) circle (1pt) node [above] {$+$}; 
    }
\foreach \n in {11}{%
        \draw[fill] (\n/2,-2) circle (1pt) node [above] {$+$};    
    }
%--------------

%-------------- Right L
\foreach \n in {1,2,...,3}{%
        \draw[fill] (\n+3,0) circle (1pt) node [above] {}; 
    }
    
\foreach \n in {1,2,...,6}{%
        \draw[fill] (\n/2+3,0) circle (1pt) node [above] {};    
    }
--------------

\node [below] at (0.5,-0.2) {$0$};

\foreach \n in {0,1,2,2.5}{%
        \draw[fill] (-\n,0) circle (1pt) node [above] {$-$}; 
    }
    
\foreach \n in {1,3}{%
        \draw[fill] (-\n/2,0) circle (1pt) node [above] {$+$};    
    }
    
%%--------------------------------------
%%--------------------------------------

\draw (-6,-2) -- (6,-2);
\node [below] at (0.5,-2.2) {$0$};
\node [below] at (-1.5,-2.2) {$-L$};

\node [below] at (-4,-2.2) {$-N$};

\node [below] at (5.5,-2.2) {$$};
%\node [below] at (-5.5,-2.2) {$-n$};

%------------- Between 0 and L
\foreach \n in {-3,...,3}{%
        \draw[fill] (\n,-2) circle (1pt); 
    }
    
\foreach \n in {-6,...,6}{%
        \draw[fill] (\n/2,-2) circle (1pt);    
    }
%--------------

%-------------- Between -N and 0
    
\foreach \n in {4,...,8}{%
        \draw[fill] (-\n/2,-2) circle (1pt) node [above] {$+$};    
    }
    \foreach \n in {9,...,10}{%
            \draw[fill] (-\n/2,-2) circle (1pt) node [above] {$+$};   
     %   \draw[fill] (-\n/2,-2) circle (1pt) node [above] {$\omega$};    
    }
%--------------

%-------------- Left -N
\foreach \n in {6}{%
        \draw[fill] (-\n,-2) circle (1pt) node [above] {$+$}; 
    }
    
    \foreach \n in {6}{%
        \draw[fill] (-\n,-2) circle (1pt) node [above] {$+$}; 
    }
    
\foreach \n in {11}{%
        \draw[fill] (-\n/2,-2) circle (1pt) node [above] {$+$};    
    }
%--------------

%-------------- Right L
\foreach \n in {1,2,...,3}{%
        \draw[fill] (\n+3,-2) circle (1pt) node [above] {}; 
    }
    
\foreach \n in {1,2,...,6}{%
        \draw[fill] (\n/2+3,-2) circle (1pt) node [above] {};    
    }
%--------------

%-------------- Node L_1

%\node [below] at (0,-0.2) {$I^*$};
%\node  at (1,0) {$($};
%\node  at (2,0) {$)$}; 

%\node [above] at (0.25,-2) {$\downarrow$}; 
 %\node[align=center, above] at (0.25,-1.5) {$+$ phase};

%\node [above] at (2.5,-2) {$\downarrow$}; 
%\node[align=center, above] at (2.5,-1.5) {$+$ phase};
 
%\node [below] at (-0.5,-2.2) {$L_1$};

\foreach \n in {0,1,2}{%
        \draw[fill] (-\n,-2) circle (1pt) node [above] {$-$}; 
    }
    
\foreach \n in {1,3,5}{%
        \draw[fill] (-\n/2,-2) circle (1pt) node [above] {$+$};    
    }
 
 \node[align=center, below] at (0,-3)%
{Figure 1 : Left $\pm$ Neighborhoods of $\omega_{\rm alt}$};
 \end{tikzpicture}
\end{center}

By (\ref{gplus}), 
%and (\ref{constrLimit}), 
for a $\mu^+$-a.s. given $\omega$, the value taken by  $g^+$ will be the infinite-volume limit of  the magnetisation of the finite-volume Gibbs measure of a Dyson-model on $\Lambda=[0,n]$, with the same decay $\alpha <2$ and $\omega$-dependent inhomogeneous external fields $h_x[\omega], x \geq 0$. For configurations $\omega : =\omega^-$ on the {sub-}neighborhood $\mathcal{N}_{N,L}^{-,{\rm left}}(\omega_{{\rm alt}})$, one gets external fields 

$$
\forall x \geq 0,\; h_x[\omega]=\sum_{k=1}^L \frac{(-1)^k}{(k+x)^\alpha} - \sum_{k=L+1}^N \frac{1}{(k+x)^\alpha} + \sum_{k\ge N} \frac{\omega_{-k}}{(k+x)^\alpha} +  \sum_{k \geq n} \frac{1}{(k+x)^\alpha}
$$

%$$
%\forall x \geq 0,\; h_x[\omega]=\sum_{k=1}^L \frac{(-1)^k}{(k+x)^\alpha} - \sum_{k=L+1}^N \frac{1}{(k+x)^\alpha} + \sum_{k=N}^n \frac{\omega_{-k}}{(k+x)^\alpha} + 2 \sum_{k \geq n+1} \frac{1}{(k+x)^\alpha}
%$$
while for $\omega:=\omega^+ \in \mathcal{N}_{N,L}^{+,{\rm left}}(\omega_{{\rm alt}})$, we get:

$$
\forall x \geq 0,\; h_x[\omega]=\sum_{k=1}^L \frac{(-1)^k}{(k+x)^\alpha} + \sum_{k= L+1}^N \frac{1}{(k+x)^\alpha} + \sum_{k\ge N} \frac{\omega_{-k}}{(k+x)^\alpha} +  \sum_{k \geq n} \frac{1}{(k+x)^\alpha}
$$

%$$
%\forall x \geq 0,\; h_x[\omega]=\sum_{k=1}^L \frac{(-1)^k}{(k+x)^\alpha} + \sum_{k=L+1}^N \frac{1}{(k+x)^\alpha} + \sum_{k=N}^n \frac{\omega_{-k}}{(k+x)^\alpha} + 2 \sum_{k \geq n+1} \frac{1}{(k+x)^\alpha}
%$$

We recognize  a  long-range RFIM  with dependent  biased, disordered external field, whose distribution is linked to the original measure $\mu$ itself via the distribution of the past. When the fields are homogeneous one can  use correlation inequalities and uniqueness via Lee-Yang \cite{YL} type arguments -- as were e.g. used to prove essential discontinuities for the decimation of Dyson model in Section 3.2. -- but here this external field will change signs, depending on  $x \in [0,n]$. For $n, L, N(L)$ large enough, it starts by being negative at  $0$  and, due to the $+$-boundary procedure far away, it becomes positive for $x$ large.

\begin{center}
 \begin{tikzpicture}[]
    
\draw (-6,0) -- (6,0);

\node [below] at (-2.5,-0.2) {$-L$};

\node [below] at (3.5,-0.2) {$n$};

\node [below] at (-5,-0.2) {$-N$};
%\node [below] at (-5,-0.2) {$-n$};

%------------- Between 0 and L
\foreach \n in {-3,...,3}{%
        \draw[fill] (\n,0) circle (1pt); 
    }
    
\foreach \n in {-6,...,6}{%
        \draw[fill] (\n/2,0) circle (1pt);    
    }
%--------------

%-------------- Between -N and 0
    
\foreach \n in {6,...,10}{%
        \draw[fill] (-\n/2,0) circle (1pt) node [above] {$-$};    
    }
%--------------

%-------------- Left -N
\foreach \n in {5}{%
        \draw[fill] (-\n,0) circle (1pt) node [above] {$+$}; 
    }
    
\foreach \n in {11}{%
        \draw[fill] (-\n/2,0) circle (1pt) node [above] {$+$};    
    }
%--------------

%-------------- Right L
\foreach \n in {1,2,...,3}{%
        \draw[fill] (\n+3,0) circle (1pt) node [above] {$+$}; 
    }
    
\foreach \n in {1,2,...,6}{%
        \draw[fill] (\n/2+3,0) circle (1pt) node [above] {$+$};    
    }
%--------------

%-------------- Node L_1

%\node [below] at (0,-0.2) {$I^*$};
%\node  at (1,0) {$($};
%\node  at (2,0) {$)$}; 

\node [above] at (0.25,0) {$\downarrow$}; 
 \node[align=center, above] at (0.25,0.5) { $h_x(\omega)<0$};

\node [above] at (2.5,0) {$\downarrow$}; 
\node[align=center, above] at (2.5,0.5) {$h_x(\omega)>0$};
 
\node [below] at (-0.5,-0.2) {$0$};

\foreach \n in {1,2}{%
        \draw[fill] (-\n,0) circle (1pt) node [above] {$-$}; 
    }
    
\foreach \n in {3,5}{%
        \draw[fill] (-\n/2,0) circle (1pt) node [above] {$+$};    
    }
    
%%--------------------------------------
%%--------------------------------------

\draw (-6,-2) -- (6,-2);

\node [below] at (-2.5,-2.2) {$-L$};

\node [below] at (3.5,-2.2) {$n$};

\node [below] at (-5,-2.2) {$-N$};

%\node [below] at (-5,-2.2) {$-n$};

%------------- Between 0 and L
\foreach \n in {-3,...,3}{%
        \draw[fill] (\n,-2) circle (1pt); 
    }
    
\foreach \n in {-6,...,6}{%
        \draw[fill] (\n/2,-2) circle (1pt);    
    }
%--------------

%-------------- Between -N and 0
    
\foreach \n in {6,...,10}{%
        \draw[fill] (-\n/2,-2) circle (1pt) node [above] {$+$};    
    }
%--------------

%-------------- Left -N
\foreach \n in {5}{%
        \draw[fill] (-\n,-2) circle (1pt) node [above] {$+$}; 
    }
    
\foreach \n in {11}{%
        \draw[fill] (-\n/2,-2) circle (1pt) node [above] {$+$};    
    }
%--------------

%-------------- Right L
\foreach \n in {1,2,...,3}{%
        \draw[fill] (\n+3,-2) circle (1pt) node [above] {$+$}; 
    }
    
\foreach \n in {1,2,...,6}{%
        \draw[fill] (\n/2+3,-2) circle (1pt) node [above] {$+$};    
    }
%--------------

%-------------- Node L_1

%\node [below] at (0,-0.2) {$I^*$};
%\node  at (1,0) {$($};
%\node  at (2,0) {$)$}; 

\node [above] at (0.25,-2) {$\downarrow$}; 
 \node[align=center, above] at (0.25,-1.5) {$h_x(\omega)>0$};

\node [above] at (2.5,-2) {$\downarrow$}; 
\node[align=center, above] at (2.5,-1.5) {$h_x(\omega)>0$};
 
\node [below] at (-0.5,-2.2) {$0$};

\foreach \n in {1,2}{%
        \draw[fill] (-\n,-2) circle (1pt) node [above] {$-$}; 
    }
    
\foreach \n in {3,5}{%
        \draw[fill] (-\n/2,-2) circle (1pt) node [above] {$+$};    
    }
 
 \node[align=center, below] at (0,-3)%
{Figure 2: Inhomogeneous $\omega$-dependent external fields};
 \end{tikzpicture}
\end{center}

On the contrary, on the neighborhood $\mathcal{N}_{N,L}^{-,{\rm left}}$, the inhomogeneous magnetic field $h_x(\omega)$ will stay negative  far enough to the past so that a $-$-phase is still felt at the origin in the  limits, while on the   neighborhood $\mathcal{N}_{N,L}^{+,{\rm left}}$, a $+$-phase is always selected for $N$ and $L$ of adjusted size.
In the last case, we need to evaluate the effect of large interval of minuses on its outside, faraway through an intermediate neutral interval, and eventually the lack of $g$-measure property is a consequence of the {\em entropic repulsion in wetting phenomena} described above. The precise and rigorous proof is more involved and delicate, so we omit it in these notes and refer to \cite{BEELN18}.

% To prove the essential discontinuity and in some sense "some" wetting beyond the origin through the alternating region, we first use the interface result of \cite{CMPR} (see also \cite{CMP17}) to state and prove in Section 3 a 
%more standard 
%wetting result that we relate to entropic repulsion.

%  shortly described now.\\

%To express {\bf that} this wetting {\bf does occur} 

  \subsection{Other results -- external fields; random b.c. and metastates}
  
 % \subsection{}
 
  \subsubsection*{\bf External fields :} 
  
 \hspace{.5cm}  A general study for inhomogeneous external field or alternated ones is still to be done. Uniqueness has been proved in various situations, in a series of papers of Kerimov (see {\em e.g.} \cite{Kerimov2007, Kerimov2009}), while Bissacot {\em et al.} have considered both uniqueness and phase transition issues in the case of decaying fields. Correlated external fields are currently studied by Littin in a work in progress.
  
  \subsubsection*{\bf Disordered fields :} 
  
  \hspace{.5cm} As for higher dimensionnal $n.n.$ Ising model where randomness yeild a dimension reduction, adding a random $i.i.d.$ magnetic field reduces the phase transition to ranges $\alpha \leq \frac{3}{2}$. Uniqueness was known by Aizenman-Wher type arguments, while a contour proof of phase transition has been provided by \cite{COP09}. For $\alpha \geq \frac{3}{2}$, the peculiarities of the unique phase according to realizations of the external fields have been described in \cite{COP12}. 
%  AW,  Cassandro {\em et al.}
  
\subsubsection*{\bf Random b.c. and metastates:} 

\hspace{.5cm} As is higher dimensional standard $n.n.$ Ising model, the behaviour under random {\em incoherent} b.c., in the sense that they are drawn from untypical b.c. (say {\em i.i.d.} when phase transition holds) also leads to a difficult toy model for spin-glasses \cite{EMN, ENS1, ENS2}. In the corks in progress  \cite{EELN19, ELN19}, we consider the Dyson model with  b.c. drawn from $i.i.d.$ sequences and describe a non-trivial metastate behaviour, with again a critical decay value $\alpha=\frac{3}{2}$ discriminating between two different global behaviours.

%***************FROM PreJSP*******************

\section{Long-range Ising model in dimension two}

\hspace{.5cm} In dimension two, let us focus on two different type of models,  an {\em isotropic} one where everybody interacts with everybody with a strength decaying with the distance (for decays $\alpha>2$), or  {\em anisotropic} models, where only sites on the same horizontal or vertical axis interact (but possibly for longer decays $\alpha >1$). We investigate the translation-invariance of extremal states, in the direction of the validity of AH theorem in the most common isotropic case, and on the other hand we describe the existence of  rigid (extremal and non translation-invariant) Dobrushin states in the anisotropic case with slow decays. 

In this section, we describe the results of \cite{CELNR18}, and add a detailed proof of van Beijeren's techniques, already known for long-range models but whose proof was well hidden in the appendix of a (not obviously) related paper of Bricmont {\em et al} \cite{BLPO79}.

\subsection{Absence of Dobrushin states in  the isotropic cases}

\hspace{.5cm} Consider classical $2d$ extensions of long-range Dyson models, with an isotropic pair potential, ({\em i.e.} a  uniform polynomial  decay $\alpha > 2$) of the form

\begin{equation}\label{coupling}
J^\alpha_{x,y} = J^{n.n.}_{x,y} + \frac{1}{|x-y|^\alpha},\; \forall x,y \in \mathbb{Z}^2
\end{equation}

In our ferromagnetic framework, phase transition at any decay $\alpha>2$ holds at low temperature by stochastic dimination of the corresponding $n.n.$-model. Nevertheless, different critical values, as in $d=1$ with $\alpha=2$, have  been identified although they do not manifest in phase transition phenomenon. In Fourier analysis techniques or mean-field/lace expansion questions,  $\alpha=4$ appear to be an important threshold, while at $\alpha \leq 3$, some peculiarities appear for non-Ferromagnetic or disordered models, and the (Gertzik)-Pirogov-Sinai picture is 'probably' not valid anymore \cite{Ger76, Ger80, PS}.

In \cite{CELNR18}, we  mainly consider decays $2 < \alpha < 4$, distinguishing between a 'medium-range' picture $3<\alpha \leq 4$, and a 'very long-range one' $2 < \alpha \leq 3$. By stochastic domination of the corresponding $n.n.$-case, phase transition holds at low temperature  $T \leq T_c(\alpha,d=2)$ and the pure phases $\mu^-$ and $\mu^+$ are built by the standard monotone weak limit procedure. 

We write $\omega=(\pm,h)$ for the so-called Dobrushin b.c.\ centered at height $h\in\Z$:
\begin{align}
\omega_x=
\left\{
\begin{matrix}
+1, & \text{ on } \{ (x_1,x_2)\in\Z^2: x_2\geq h\}\\
-1, & \text{ on } \{ (x_1,x_2)\in\Z^2: x_2 < h\}
\end{matrix}
\right.
\end{align}

For a given height $h$, write $\mu^{(\pm,h)}$ for 
any (sub-sequential) weak limit of sequences ${(\mu^{(\pm,h)}_{\Lambda})}_{\Lambda}$. As in previous sections,  the Gibbs measure  $\mu^{(\pm,h)}$ is called a Dobrushin state if it is extremal and is \emph{not} translation-invariant. In this work, we exclude their existence in both cases;  we either use an energy estimate and, as in Section 3.2 and 3.3, 'Equivalence of b.c.' from \cite{BLP79} in the shorter-range case $0<\alpha<3$, or a strategy of Fr\"ohlich/Pfister \cite{FrPf81} using relative entropy estimates  to exclude cohabitation of  translation-invariance and extremality in the longer range cases $2 < \alpha \leq  3$.

\subsubsection*{\bf Medium ranges $3<\alpha \leq 4$:} 

\hspace{.5cm} In this case, ferromagnetism is not needed and the results got are more general. By comparing Hamiltonians of different Dobrushin b.c. located at two consecutive planes,  we see that this energy difference is already uniformly bounded for decays $\alpha>3$, allowing us to avoid entropic considerations, while for longer-range decays we shall see see that relative entropy estimates and ferromagnetism are needed to incorporate entropic effects. 
%In order to emphasize the short-range '$n.n.$-like' character for medium-range polynomial decays $\alpha >3$, we provide energy estimates between 
%\arno{
%the states obtained  by $\pm$ Dobrushin b.c. and their translates,
%}
% which allow  us to conclude, as in {\em e.g.} \cite{DS85}, that for those ranges at least, the  states obtained via Dobrushin boundary conditions  are translation-invariant. 
%\arno{This will  shorten the proof of some interface fluctuation results in this case. Similar estimates have allowed to really treat these models as 
%short-ranges ones, with Peierls contours and estimates etc. (see ???).} 
%{\bf 

The energetic  observation we use is that the difference between two Dobrushin conditions is obtained by flipping all spins in two half-lines, so if the maximal energy between a half-line left of the origin and a half-plane right of the origin  is uniformly bounded, the arguments of equivalence of boundary conditions of \cite{BLP79} apply and we can conclude that there is no 'pure' interface Gibbs state, or said differently no interface state.  What we show in \cite{CELNR18} is that it holds for decays $\alpha>3$.

%We remark as an aside that this argument does not need the ferromagnetic character of the model.
%} 

%\begin{description}
%\item{

%{\bf Energy difference in a finite box with Dobrushin boundary conditions}}

Denote $H^{\pm}_{\Lambda}(\sigma)$ the Hamiltonian with Dobrushin b.c. in $\Lambda=\Lambda_L=\big( [-L,+L] \cap \mathbb{Z} \big)^2$ :

\[
- H^{\pm}_{\Lambda}(\sigma) = \sum_{x,y\in \Lambda} \sigma_x \sigma_y J_{x,y}+ \sum_{x\in \Lambda, y\in \Lambda
^u}\sigma_x  J_{x,y} - \sum_{x\in \Lambda, y\in \Lambda^d}\sigma_x  J_{x,y}
\]
where  
%\arno{
$\Lambda^u=\{ (x,y): y\geq 0\} \cap \Lambda^c$ and $\Lambda^d=\{ (x,y): y< 0\} \cap \Lambda^c$.
%} 
Let $\overline{H}^{\pm}_{\Lambda}(\sigma)$ be defined as the shifted Hamiltonian,  with upward-shifted Dobrushin b.c.:
\[
- \overline{H}^{\pm}_{\Lambda}(\sigma) = \sum_{x,y\in \Lambda} \sigma_x \sigma_y J_{x,y}+ \sum_{x\in \Lambda, y\in \Lambda^{u+1}}\sigma_x  J_{x,y} - \sum_{x\in \Lambda, y\in \Lambda^{d-1}}\sigma_x  J_{x,y}
\]
where 
%\arno{
$\Lambda^{u+1}=\{ (x,y): j\geq 1\} \cap \Lambda^c$ and $\Lambda^{d-1}=\{ (x,y): j\leq 0\} \cap \Lambda^c$. Then we can estimate
 %first assuming $\alpha > 3$

\[
\begin{split}
& |H^{\pm}_{\Lambda}(\sigma)- \overline{H}^{\pm}_{\Lambda}(\sigma)| \\
 &= \left |\sum_{x\in \Lambda, y\in \Lambda^u}\sigma_x  J_{x,y} - \sum_{x\in \Lambda, y\in \Lambda^{u+1}}\sigma_x  J_{x,y}+  \sum_{x\in \Lambda, y\in \Lambda^{d-1}}\sigma_x  J_{x,y} - \sum_{x\in \Lambda, y\in \Lambda^d}\sigma_x  J_{x,y} \right | \\
& \leq  \sum_{(y_1,0) \in \Lambda^c}  \sum_{(x_1,y_1)\in \Lambda} O(|(x_1-y_1)^2+y_1^2 |)^{-\alpha/2}\\
%& =  \sum_{(y_1,0) \in \Lambda^c}  \sum_{x=-L}^L \sum_{j_x=-L}^L O(|(i_x-i_y)^2+j_x^2 |^{-\alpha/2}) \\
%& \leq   \sum_{(i_y,0) \in \Lambda^c}  \sum_{i_x=-L}^L \sum_{j_x=-\infty}^{\infty} O(|(i_x-i_y)^2+j_x^2 |%^{-\alpha/2})\\
%& =  \sum_{(i_y,0) \in \Lambda^c}  \sum_{x_1=-L}^L  O(|x_1-y_1|^{1-\alpha}) \\
& \leq \sum_{y_1=L+1}^{\infty}  \sum_{x_1=0}^L O\left ( (y_1-x_1)^{1-\alpha} + (x_1+y_1)^{1-\alpha}\right ) 
\end{split}
\]

so that|$H^{\pm}_{\Lambda}(\sigma)- \overline{H}^{\pm}_{\Lambda}(\sigma)| \leq C(\alpha)=O(L^{3-\alpha})$
 which is uniformly bounded for $\alpha > 3$. 

Now, one proceed as \cite{BLP79} by using 'Equivalence of boundary conditions" : Finite energy difference implies that the states obtained as weak limits
 are absolutely continuous  w.r.t. each other and should have the same components in their extremal decomposition. When the limit state is an extremal Gibbs measure, the state and its translate would thus be equal, and thus the state would be translation-invariant.
%}).

As described briefly in \cite{CELNR18}, the case of fast decays $\alpha >3$ falls in fact  within the framework of the Gertzik-Pirogov-Sinai theory \cite{PS}. These models satisfy a Peierls condition at low enough temperature as shown in \cite{Ger76, Ger80}. In such a framework, all the Gibbs measures should be translation-invariant, as described in the review \cite{DS85}. From this, coupled with the fact recently extended to more general contexts  by Raoufi  \cite{Raoufi18} that the $\mu^+$ and $\mu^-$ states are the only translation-invariant extremal states, one gets also the convex decompositions in terms of these pure states.
For the standard Dobrushin b.c. located at the origin, one recovers

 $$\mu^{(\pm,0)}=\lim_{\Lambda\uparrow\Z^2}\mu_{\Lambda}^{(\pm,0)}=\frac{1}{2} (\mu^- +  \mu^+)$$

\subsubsection*{\bf Longer ranges $2<\alpha<3$:} 

\hspace{.5cm} In this case, we first need to consider the zero-temperature case and investigate the asymptotic behavior of the { \em Energy difference for Dobrushin ground states and shifted ground state}, obtained  by shifting the spin on a half-line only. Indeed, although the maximal interaction energy between a half-line left of the origin and a half-plane right of it is infinite, we show in \cite{CELNR18} that the expected interaction energy in a state with Dobrushin boundary conditions still remains finite. We use there both the ``antisymmetry'' between up and down and the ferromagnetic character of the interaction. The argument uses the fact that the interaction of the negative 
half-line 
$\{ i < 0, j=0 \}$ and the positive half-line $\{ i \geq 0, j=0 \}$ is finite, while the interaction of the half-line with any plus spin above the line is canceled by the interaction with the reflected minus spin below the line. %Indeed, the argument breaks down if the interaction would for example have  alternating signs in the vertical direction. 

To see this, split the lattice $\mathbb{Z}^2$ into $A^+=\{ (i,j): j\geq 1 \}\cup \{ (i,0): i> 0\}$, $A^-=\{ (i,j): j\leq -1 \}$ and $A^0=\{ (i,0): i\leq0 \}$. Consider the Dobrushin 
ground states $\sigma_{GS}$ in the sense that we put all $+1$ in $A^+\cup A^0$ and $-1$ in $A^-$ with energy $H^{\pm}(\sigma_{GS})$. We call after $\sigma_{GS,step}$ the configuration $\sigma_{GS}$ which is flipped on the half line $A^0$, 
consisting thus in $+1$ in $A^+$ and $-1$ in $A^0\cup A^-$, and estimate the energy difference. Then
\[
-H(\sigma_{GS}) = \frac12\sum_{x,y\in A^+} J_{xy} + \frac12\sum_{x,y \in A^-}J_{xy} - \sum_{x\in A^+, y\in A^-} J_{xy} + \sum_{x\in A^+, y\in A^0}  J_{xy} -\sum_{x\in A^0, y\in A^-} J_{xy}
\]
and
\[
-H(\sigma_{GS,step}) = \frac12\sum_{x,y\in A^+} J_{xy} + \frac12\sum_{x,y \in A^-} J_{xy} - \sum_{x\in A^+, y\in A^-} J_{xy}{-} \sum_{x\in A^+, y\in A^0}  J_{xy} {+}\sum_{x\in A^0, y\in A^-} J_{xy}
\]
writing as before $x=(x_1,x_2), y=(y_1,y_2)$, the energy difference is equal to
\[
\begin{split}
\big|H(\sigma_{GS}) - H(\sigma_{GS,step})\big|& = 2 \left| \sum_{x\in A^+, y\in A^0}  J_{xy} -  \sum_{y \in A^0, x\in A^-} J_{xy}\right| \\
%& = \left | \sum_{i_y = - \infty}^0 \sum_{i_x\in \mathbb{Z}} \sum_{j_x=1}^{\infty} J_{xy} + \sum_{i_y=-\infty}^0 \sum_{i_x=1}^{\infty} J_{xy} - \sum_{i_y = - \infty}^0 \sum_{i_x\in \mathbb{Z}} \sum_{j_x=-\infty}^{-1} J_{xy}\right |.
\end{split}
\]
which by symmetry of the couplings $J_{xy}$ is uniformly bounded for $\alpha >2$.

%\subsubsection{Consequence for Dobrushin states} Prove also T.I., using ferromagnetism and energy estimate above.

%We should more follow \cite{Pfis81} or adapt long-range treatment of \cite{FrPf81}, because \cite{DS85} relies on Pirogov-Sinai techniques reserved to finite range periodic hamiltonians satisfying Peierls bounds.)******
%...It follows the line of an argument of Pfister \cite{Pfis81} after a physical intuitive arguments of Herring and Kettel \cite{HK51} (and maybe also \cite{FrPf81} or \cite{Shlo80} or \cite{Sakai75}.

%\arno{From this, the argument of \cite{DS85} also sketched somewhere by Aernout, also described in by Pfister (Fr\"ohlich ? Kunz ?), provides equivalence of boundary conditions and also translation invariance if i remember well. I did not find the sketch, it should be derived a bit, it  uses integration of the difference energy w.r.t. the plus state.}
%\item{

%{\bf  Positive Temperatures}
%\arno{From Aernout's mail : TO BE INCORPORATED }
A similar argument will still hold at low but positive temperatures, and we sketch it now. For a complete rigorous proof, consult \cite{CELNR18}. The main observation is that the interaction energy of a spin interacting with a half-plane  at distance $l$ is maximally of order $O(l^{2- \alpha}$), but its expectation in the Gibbs state with Dobrushin b.c. $(\pm, h)$ at more or less the same height is $O(l^{1- \alpha})$.
Summing over the line just above the  interface gives then that the total expected energy cost of shifting is uniformly bounded, thus the relative entropy, which between two Gibbs measures corresponds to the  expectation of Hamiltonians difference, computed in one of them  between the two putative Dobrushin states is finite. This implies that, once they are extremal, these states are the same, using  again the same relative entropy arguments as {\em e.g.} in \cite{BLP79,FrPf81}. This implies the translation invariance of the measures got by weak limits of Dobrushin b.c. and the absence of Dobrushin {\em states}.

%To see this, take {\em e.g.} $k=1$ for comparing Dobrushin b.c. which are shifted by distance 1 and
 %let $\tau(x,y)$ be a boundary spin  $(|x| > L)$ and $\sigma(x,y)$ a spin in the strip $-L < x < L$. The ``strip Gibbs measure'' is defined on the infinite strip with width $2L-1$,  with Dobrushin boundary conditions $\tau_{x,y}=+$ when $y \geq 1$, and $\tau_{x,y} = -$ when $y \leq 0$.
%Call it $\mu_{L,1}$; it  has the property (antireflection symmetry)
%that
%$\mu_{L,1} (\sigma_{x,y} = - \mu_{L,1} (\sigma_{x, 1-y})$. Consider $\tau_{L,1}$ a boundary spin on the line $y=1$ which to flip requires to shift the boundary condition. Assume $|||$ is reasonably larger than $|L|$ so the distance to the boundary is $|l-L|$.

%The expected interaction of this boundary spin with the sites in the strip is a sum of terms
%$\mu(\tau_{l,1} \sigma(x,y) ){\max\{|l-x)|,  |y-1| \}}^{-\alpha}$.
%now write  the sum as a sum over  pairs $(x,y), (x,1-y)$, with  all $x$ in the strip, and all $y \geq 1$.
%Then one sees that the expected interaction of this spin is of order $O(|l-L|^{2-1-\alpha})=O(|l-L|^{1 -\alpha})$,
%summing over $l$ gives a bounded sum when $\alpha$ is larger than 2. This gives $\mu_{L,1}  (\Delta H)$ so that the relative entropy between $\mu_{L,1}$ and $\mu_{L,0}$ is of the form $\mu_{L,1} (\Delta H)$, which  remains  uniformly bounded as $L$ grows to infinity.

%\end{description}

 %JSP ...PreJSP hanuary 2018**************\\
 %\\
 
 \subsection{Rigidity in anisotropic cases}
 
% ***********************************

%{\bf Anisotropic long-range/$n.n.$ Ising models in dimension 2}
%\item{{\bf Example II - anisotropic long-range/$n.n.$ Ising models}}

\hspace{.5cm} We consider two different but simlar anisotropic long-range models on $\Z^2$ here : First, a mixed long and $n.n.$ translation-invariant interaction whose interaction  are  $n.n.$ vertically and 'Dyson-like' horizontally, {\em i.e.} 
of the form 

%%\begin{eqnarray*}\label{mixcoupling}
$$
J_{x,y} = 1 \; \; \; \; {\rm if} \; \; x_1=y_1  \;\; {\rm and} \; \; \; \; y_2=x_2 \pm 1;\;\; 
 J_{x,y} = 0 \; \; \; \; {\rm if} \; \; \; \; |y_2-x_2| > 1
$$
and
$$
J_{x,y} = |x_1-y_1|^{-\alpha_1} \; \; \; \; \;  {\rm if} \; \; \; \; x_2=y_2\\
$$

%$$\delta(x,0)\delta (|y|,1) + \delta(y,0) |x|^{- \alpha} $$
%with {\bf $1< \alpha \leq 2$}.

%\begin{itemize}

%\item 
%Similarly to the next subsection, the extension of the Van Beijeren's proof by Bricmont {\em et al.} in the Appendix of  \cite{BLPO79} applies,  as it can be extended to long-range models, 
%which applies
%seems to be true 
%as long as some symmetries are kept. 
%It gives horizontal-interface states in $d=2$, with standard Dobrushin b.c.\\
%In this case, however,  also van Beijeren's original proof could  be used \cite{vB}, see Section 4.2.  
%{\bf ??Sketch of the rigidity proof of \cite{BLPO79} for $1 < \alpha_1 \leq 2$ :Next subsection????}

%{\bf ?? Next Question: Impose Dobrushin boundary conditions in the wrong direction, trying unsuccessfully to make vertical interfaces. There, could we get again fluctuations ?  Do you get anomalous fluctuations- like Dynkin78son interface in $d=1$ eg- for $\alpha_ \in (1,2)$ ? What about an invariance principle result with a L\'evy bridge ?!!!} 

%\end{itemize}

%\item{

%{\bf Example -- Biaxial, possibly anisotropic, long-range models}

and secondly a  'Dyson-like' long-range interactions in both horizontal and vertical directions, with not necessarily the same powers, $\alpha_1,\alpha_2$,  where at east one of the two is in $(1,2)$:
%or in $(2,\infty)$.

$$
J_{x,y} =|x_2-y_2|^{-\alpha_2} \; \; \; \; \;  {\rm if} \; \; \; \; x_1=y_1
,\;
J_{x,y} = |x_1-y_1|^{-\alpha_1} \; \; \; \; \;  {\rm if} \; \; \; \; x_2=y_2
$$

%This is a particular case of the models treated in the appendix of \cite{BLPO79}; their arguments imply rigidity of the interface states for these long-range models in dimension 2, either in one or two directions depending on whether $\alpha_1, \alpha_2$, or both, are between 1 and 2. 

%{\bf ?? (Check the long-range adaptation -- Phase Transition in both 'dimensions' there in addition to the long-range caracter)??}

%\end{itemize}
%****************PreJSP hanuary 2018****************

%****************************************************

Let us see, as proposed in \cite{vB} and described in the appendix of \cite{BLPO79}, that in dimension 2 it is  possible to get rigidity  for anisotropic, but still rather symmetric long-range models. 

In fact, this is the case  as soon as one keeps :

- Some monotonicity properties of the couplings  $J$ as a function of  the graph distance.

- Symmetry w.r.t. the horizontal axis $x_2=0$ (the one of the Dobrushin b.c.).

- Spontaneous magnetization of the one-dimensional system with the same coupling (decoupled from the rest of the lattice).

We provide a description for the more general model. Recall that it has interactions along the horizontal  and vertical axis only, with polynomial decays $\alpha_1$ and $\alpha_2$ :
\begin{align}
J_{xy} &=|x_2-y_2|^{-\alpha_2} \quad  \text{ if }   x_1=y_1\nonumber\\
J_{xy} &= |x_1-y_1|^{-\alpha_1} \quad  \text{ if }  x_2=y_2\nonumber\\
J_{xy}&=0 \quad \text{ otherwise}
\end{align}
\begin{theorem}
For the anisotoropic models described above, at low enough temperature and slow horizontal decay $1<\alpha_1\leq2$, there exists a Dobrushin state $\mu^\pm$, non-translation-invariant and extremal, selected as weaks limit with horizontal boundary conditions $(\pm,0)$. It is  such that on the horizontal line $\Delta_1:x_2=0$,
$$
\forall x=(x_1,0) \in \Delta_1,\; \langle \sigma_x \rangle^\pm > 0
$$
\end{theorem}

{\bf Remark :}  The vertical decay does  not play any role  with this horizontal Dobrushin b.c. but might enters in fluctuations in the case when the interface is not rigid, but rough.

Our proof is a detailed  adaptation to the case of  infinite-range models performed in the Appendix $B$ of \cite{BLPO79}. The particular form of the interactions, with pair-potentials along two lines only, allows us to present completely the proof and to take some shortcuts. Mathematically speaking, it also comes as a generalization already noticed by van Beijeren \cite{vB} in its proof of Dorushin's rigidity result in dimension 3 \cite{Dob72}. To get strict positivity of the magnetization under the putative Dobrushin state, and thus non-translation invariance of any weak limit, one compares it with the spontaneous magnetization of a one-dimensional 'Dyson-like' auxiliary system  using a duplicate trick as follows   :

\begin{enumerate}
\item {\bf Step 1:  Duplicate} a configuration $\sigma \in \{-1,+1\}^{\mathbb{Z}^2}$ from the original $2d$-system with Hamiltonian $H^\pm$ of a $1d$ long-range Ising model with the same polynomial decay $\alpha_1$,  restricted to the horizontal line, decoupled from the rest of the plane, with $+$-b.c. instead of $\pm$. Call it $H'^+$, write its ferromagnetic coupling $J'$ and pick a configuration $\sigma'$ according to it. By independent duplication, write formally the joint Hamiltonian
\be\label{JointH}
{\tt H}^{(\pm,+)}(\sigma,\sigma') : = H^\pm(\sigma) + H'^+(\sigma')
\ee
Consider a volume $\Lambda=\Lambda_L =\big( [-L,L] \cap \Z \big)^2$ and partition it naturally as $\Lambda = \Lambda^+ \cup \Lambda^0 \cup \Lambda^-$, where  $\Lambda^0$ is the line  $\{x_2=0\}$, $\Lambda^+$ the (strict) upper half-plane $\{x_2>0\}$ and $\Lambda^-$ the lower one $\{x_2 < 0 \}$. For these volumes, (\ref{JointH}) reads
\begin{eqnarray}
- {\tt H}^{(\pm,+)}_{\Lambda,\Lambda^0} (\sigma,\sigma') = \sum_{x,y \in \Lambda} J_{xy} \sigma_x \sigma_y + \sum_{x \in \Lambda} \Big(\sum_{y \in \Lambda^c, y_1 \geq 0} J_{xy} \sigma_x -  \sum_{y \in \Lambda^c, y_1 < 0} J_{xy} \sigma_x\Big)  \nonumber \\
+  \sum_{x,y \in \Lambda_0} J'_{xy} \sigma'_x \sigma'_y + \sum_{x \in \Lambda_0, y \in \Lambda^c, y_1=0} J'_{xy} \sigma'_x
\end{eqnarray}
To make use of the symmetry w.r.t the horizontal axis, we define the symmetric of $x \in \Z^2$ as $\bar{x}=(x_1,-x_2)$ for any $x=(x_1,x_2) \in \Z^2$ and remark  that
\be \label{positcoupl}
J_{\bar{x}\bar{y}} = J_{xy} \geq 0 , \; J_{x,\bar{y}}=J_{\bar{x}y} \geq 0,\; J_{xy} \geq J_{x \bar{y}}
\ee

Rewrite the joint Hamiltonian $- {\tt H}^{(\pm,+)}_{\Lambda,\Lambda^0} (\sigma,\sigma')$ as
\begin{eqnarray}
& &  \sum_{x \in \Lambda^+} \Big(\sum_{y \in \Lambda } J_{xy} \sigma_x \sigma_y  + \sum_{y :  y_1=x_1, |y_2| > L} {\rm sgn} (y_2) J_{xy}\sigma_x 
 + \sum_{y : |y_1|>L, y_2=x_2} J_{xy} \sigma_x   \Big) \label{first}\\
&+& \sum_{x \in \Lambda^-} \Big( \sum_{y \in \Lambda} J_{xy} \sigma_x \sigma_y  + \sum_{y :  y_1=x_1, |y_2| > L} {\rm sgn} (y_2) J_{xy}\sigma_x 
 - \sum_{y :  |y_1|>L, y_2=x_2} J_{xy} \sigma_x   \Big) \label{second}\\ 
 &+& \sum_{x \in \Lambda^0} \Big( \sum_{y \in \Lambda^0} J_{xy} \sigma_x \sigma_y  + \sum_{y :  y_1=x_1, |y_2| < L} J_{xy}\sigma_x \sigma_y 
 + \sum_{y :  y_1=x_1, |y_2| > L} {\rm sgn} (y_2) J_{xy}\sigma_x 
\label{third}\\
& +&\sum_{y : |y_1|>L, y_2=0} J_{xy} \sigma_x  +  \sum_{y \in \Lambda_0} J'_{xy} \sigma'_x \sigma'_y + \sum_{y : |y_1|>L, y_2=0} J'_{xy} \sigma'_x
% + \sum_{y :  |y_2 | < L, y_1=0} {\rm sgn} (y_2) J_{xy}\sigma_x
 \Big)
\end{eqnarray} 

Any $x \in \Lambda^+$ can be mapped one-to-one into $\bar{x} \in \Lambda^-$ so that (\ref{second}) becomes
$$
\sum_{x \in \Lambda^+} \Big( \sum_{y \in \Lambda  } J_{\bar{x}y} \sigma_{\bar{x}} \sigma_y  + \sum_{y :  y_1=\bar{x}_1, |y_2| > L} {\rm sgn} (y_2) J_{xy}\sigma_{\bar{x}}
 - \sum_{y :  |y_1|>L, y_2=\bar{x}_2} J_{\bar{x} y} \sigma_{\bar{x}}   \Big) 
$$

while the restriction to the horizontal line (\ref{third}) can be written
\be \label{third-b}
\sum_{x \in \Lambda^0} \Big( \sum_{y \in \Lambda^0} J_{xy} \sigma_x \sigma_y + \sum_{y_1=x_1 :0< y_2 < L } \big( J_{xy} \sigma_x \sigma_y + J_{x\bar{y}} \sigma_x \sigma_{\bar{y}} \big) + \sum_{y:y_1=x_1, y_2 >L}  \big( J_{xy} - J_{x\bar{y}} \big) \sigma_x \Big)
\ee
%\medskip
\item{\bf Step 2. Use the symmetries.} By considering also sites $y$ in the upper half-plane $\Lambda^+$ and using symmetric sites, 
(\ref{first}) and (\ref{second}) merge into a term
\begin{eqnarray}
{\tt H}_{\Lambda^+,\partial \Lambda_0} &:= &\sum_{x \in \Lambda^+} \Big( \sum_{y \in \Lambda^+:y_1=x_1} \big( J_{xy} \sigma_x \sigma_y + J_{x \bar{y}} \sigma_x \sigma_{\bar{y}} \big) + \sum_{y: y_1=x_1, y_2>L} \big( J_{xy} \sigma_x - J_{x\bar{y}} \sigma_x \big) \nonumber \\
&+& \sum_{y \in \Lambda^+, y_2=x_2}J_{xy} \sigma_x \sigma_y  + \sum_{y \in \Lambda^c,y_2=x_2} J_{xy} \sigma_x  \nonumber \\
&+&  \sum_{y \in \Lambda^+:y_1=x_1} \big( J_{\bar{x}y} \sigma_{\bar{x}} \sigma_y + J_{\bar{x} \bar{y}} \sigma_{\bar{x}}\sigma_{\bar{y}} \big) + \sum_{y: y_1=x_1 y_2>L} \big( J_{\bar{x}y} \sigma_{\bar{x}} - J_{\bar{x}\bar{y}} \sigma_{\bar{x}} \big) \nonumber \\
&+& \sum_{y \in \Lambda^+, y_2=x_2} J_{\bar{x}\bar{y}} \sigma_{\bar{x}} \sigma_{\bar{y}}   - \sum_{y \in \Lambda^c,y_2=x_2} J_{\bar{x}\bar{y}} \sigma_{\bar{x}}  \Big) \nonumber
\end{eqnarray}

Use first the symmetries $J_{\bar{x} \bar{y}}=J_{xy}$ and $J_{\bar{x}y}=J_{x\bar{y}}$ for all $x,y \in \Lambda$ :
\begin{eqnarray}
{\tt H}_{\Lambda^+,\partial \Lambda_0} &= &\sum_{x \in \Lambda^+} \Big( \sum_{y \in \Lambda^+:y_1=x_1} J_{xy} (\sigma_x \sigma_y + \sigma_{\bar{x}} \sigma_{\bar{y}} ) + \sum_{y: y_1=x_1 y_2>L} (J_{xy} - J_{x\bar{y}}) \sigma_x  \nonumber \\
&+& \sum_{y \in \Lambda^+, y_2=x_2}J_{xy} \sigma_x \sigma_y  + \sum_{y \in \Lambda^c,y_2=x_2} J_{xy} \sigma_x  \nonumber \\
&+&  \sum_{y \in \Lambda^+:y_1=x_1} \big( J_{x\bar{y}} \sigma_{\bar{x}} \sigma_y + J_{x y} \sigma_{\bar{x}}\sigma_{\bar{y}} \big) - \sum_{y: y_1=x_1 y_2>L} (J_{xy}  - J_{x\bar{y}}) \sigma_{\bar{x}} \nonumber \\
&+& \sum_{y \in \Lambda^+, y_2=x_2} J_{xy} \sigma_{\bar{x}} \sigma_{\bar{y}}   - \sum_{y \in \Lambda^c,y_2=x_2} J_{xy} \sigma_{\bar{x}}  \Big) \nonumber
\end{eqnarray}
Rearrange terms to eventually get
\begin{eqnarray}
{\tt H}_{\Lambda^+,\partial \Lambda_0} &= &\sum_{x \in \Lambda^+} \Big( \sum_{y \in \Lambda^+:y_1=x_1} \big( J_{xy} (\sigma_x \sigma_y + \sigma_{\bar{x}} \sigma_{\bar{y}})
+  J_{x \bar{y}} (\sigma_x \sigma_{\bar{y}} +  \sigma_{\bar{x}} \sigma_{y}) \big) \nonumber \\
& +&    \sum_{y: y_1=x_1, y_2>L} (J_{xy} - J_{x\bar{y}}) (\sigma_x -  \sigma_{\bar{x}}) \nonumber \\
&+& \sum_{y \in \Lambda^+, y_2=x_2}  J_{xy} (\sigma_x \sigma_y + \sigma_{\bar{x}} \sigma_{\bar{y}})
+  \sum_{y \in \Lambda^c,y_2=x_2} J_{xy} (\sigma_x - \sigma_{\bar x} ) \Big)\nonumber  
\end{eqnarray}

Similar arrangements hold for the horizontal part, using also with that $J'=J$ on $\Lambda^0$, and the fact that on $\Delta^0$ our symmetry reduces to identity :
\begin{eqnarray*}
{\tt H}_{\Lambda^0,\partial \Lambda^+} &=& \sum_{x \in \Lambda^0}\Big( \sum_{y \in \Lambda^0} J_{xy} \big( \sigma_x \sigma_y + \sigma'_x \sigma'_y \big) + \sum_{y: y_1=x_1, y_2 > L}  J_{xy}\sigma_x(\sigma_y +\sigma_{\bar{y}})  \nonumber \\
% - J_{x \bar{y}} \big) \sigma_x \sum_{y: y_1=x_1 y_2>L} J_{xy} (\sigma_x - \sigma'_{x} )\\
&+& \sum_{y: |y_1|>L, y_2=0}  J_{xy} (\sigma_x + \sigma'_{x} ) + \sum_{y: y_1=x_1, 0<y_2<L} \big( J_{xy} - J_{x \bar{y}} \big) \sigma_x  \Big)  \label{approx1}
\end{eqnarray*}
Thus, we need to get information on 
$
{\tt H}_\Lambda^{\pm,+}(\sigma,\sigma')={\tt H}_{\Lambda^+,\partial \Lambda_0} + {\tt H}_{\Lambda^0,\partial \Lambda^+}.
$
We remark that it looks like a ferromagnetic system with the variable $\sigma_x \sigma_y + \sigma_{\bar{x}} \sigma_{\bar{y}}  $ and  $ \sigma_x -  \sigma_{\bar{x}} $ instead of pair-potential (quadratic) part and a  (linear) self-interaction. This is exactly the case in the trick of Percus \cite{Percus75}, as follows.
\medskip

\item{\bf Step 3. Change the variables} $(\sigma,\sigma')$ into an adaptation of the duplicate set $\{s,t\}$ of Percus  \cite{Leb74,Percus75}. It uses the symmetric $\bar{x}$ of any site $x \in \Z^2$ w.r.t $\{x_2=0\}$ 
(Compare \cite{BLPO79} page 19): 
\begin{eqnarray*}
\forall x \in \Lambda^+,\; s_x = \sigma_x + \sigma_{\bar{x}} &, & \; t_x=\sigma_x - \sigma_{\bar{x}} \nonumber \\
\forall x \in \Lambda^0,\; s_x = \sigma_x + \sigma'_x  &, & \; t_x=\sigma_x - \sigma'_x 
\end{eqnarray*}
The new variables take value in $\{-1,0,+1\}$ with some trivial constraints, but they have nice extra properties to deal with. In particular one has
\begin{eqnarray*}
\forall x,y \in \Lambda^+,\;\sigma_x \sigma_y + \sigma_{\bar{x}} \sigma_{\bar{y}}= \frac{1}{2}\big(s_x s_y + t_x t_y\big)  &, & \;  \sigma_x \sigma_{\bar{y}} + \sigma_{\bar{x}}\sigma_y  =  \frac{1}{2}\big(s_x s_y - t_x t_y\big)   \\
\forall x,y  \in \Lambda^0,\;\sigma_x \sigma_y + \sigma'_{x} \sigma'_y= \frac{1}{2}\big(s_x s_y + t_x t_y\big)  &, & \; \sigma_x \sigma'_y - \sigma'_x \sigma_y =  \frac{1}{2}\big(s_x s_y - t_x t_y \big)  \\
\end{eqnarray*}
and also among other useful relations valid for any $x$ and $y$,
$$
\sigma_x \sigma_y + \sigma_x \sigma_{\bar{y}} = s_y \frac{s_x + t_x}{2}
$$
 so that we eventually get the joint Hamiltonian
\begin{eqnarray} \label{Hst}
& & -{\tt H}^{\pm,+}_{\Lambda,\Lambda^0}(s,t) = \sum_{x \in \Lambda^+} \Big( \sum_{y \in \Lambda^+, y_1=x_1} \big( \frac{J_{xy}+J_{x \bar{y}}}{2}  s_x s_y + \frac{J_{xy} - J_{x \bar{y}}}{2}  t_x t_y \big)  \nonumber \\
&+ &\sum_{y_1=x_1, y_2 >L} (J_{xy} - J_{x \bar{y}}) t_x + \sum_{y \in \Lambda^+, y_2=x_2} \frac{J_{xy}}{2} ( s_x s_y + t_x t_y) + \sum_{y \in \Lambda^c, y_2=x_2} J_{xy} t_x  \nonumber \\
&+& \sum_{y \in \Lambda^+, y_1= x_1} (J_{xy} + J_{x \bar{y}}) \frac{s_x + t_x}{2} s _y  + \sum_{y:y_1=x_1, y_2>L} (J_{xy}-J_{x \bar{y}}) s_y \Big) \nonumber \\
&+& \sum_{x \in \Lambda^0} \Big( \sum_{y \in \Lambda^0}  J_{xy} (s_x s_y + t_x t_y) + \sum_{y_1=x_1, y_2>L} J_{xy} s_y \frac{s_x + t_x}{2}  + \sum_{y:|y_1| >L, y_2=0} J_{xy} s_y  \nonumber \\
&+& \sum_{y:y_1=x_1, 0<y_2<L} \big(J_{xy} - J_{x \bar{y}} \big)\frac{s_x + t_x}{2} \Big) \nonumber 
\end{eqnarray}

\item{\bf Step 4 : Correlation inequalities and symmetries}

For both $x,y \in \Lambda^+$, one has $J_{xy} \geq J_{x\bar{y}}$ so that from the form (\ref{Hst})  given above, it is now obvious that the joint Hamiltonian ${\tt H}$  has ferromagnetic pair interactions, or single-site interactions. Then use a generalisation of the GHS inequalities as given in the original Griffiths, Hurst and Sherman or Kelly and Sherman papers \cite{GHS70,KS68}, or the extension of them by Lebowitz \cite{Leb72,Leb74}\footnote{See also Georgii \cite{Geo88} p 447 for a precise genealogy, pair-interaction is essential.} to conclude as in \cite{vB,BLPO79} that  the coordinates are positively magnetized; in particular
$$
\langle t_x \rangle^{\pm,+}_{\Lambda^+,\Lambda^0} \geq 0.
$$
Restriction to the horizontal layer gives that
$$
\forall x \in \Lambda^0,\; \langle{\sigma_x}\rangle_{\Lambda}^{\pm} \geq \langle {\sigma'_x}\rangle_{\Lambda^0}^{+}>0 \; {\rm at \; low \; T}.
$$
The second expectation is performed for the one-dimensional Gibbs states with $+$ b.c. at the same temperature. Thus, as soon as spontaneous magnetization occurs for the latter,  this implies the existence of a non-translation-invariant Gibbs states in dimension two. 
notice that this lower-dimensional phase-transition condition is not fulfilled in the isotropic long-range models treated above, because their well-definedness requires $\alpha >2$, for which there is no phase transition in dimension one. To get such a phase transition and positive magnetization, one has to consider very long-ranges  with decays $1< \alpha_1 \leq 2$, acting on a horizontal line only. 
\end{enumerate}

%\begin{figure}[ht]
%		\centering
%		\includegraphics[width=0.33\textwidth]{SUNPOO27.JPG}
%	\caption{test}
%		\label{fig-original}
%\end{figure}

%\section{Conclusions and perspectives}
 
{\bf Acknowledgments:} I particularly thank Aernout van Enter, Pierre Picco and Yvan Velenik for particularly interesting discussions and encouragements to work on these long-range Ising models. My research during this period have been partly funded by {\em Labex B\'ezout}, funded by ANR, reference ANR-10-LABX-58. A major part of these notes have been prepared during the academic year 2017-2018 at Eurandom (TU/e, Eindhoven), supported by CNRS, Eurandom and  the dutch consortium {\em NETWORKS} (Eurandom-Lorentz Center-CWI). 

%\newpage

\end{document}